\documentclass[useAMS,usegraphicx]{mn2e}
\usepackage{amssymb}
\usepackage{lscape}
\usepackage{times}
\usepackage{pifont}
\usepackage{amsfonts}
\usepackage{amsmath}
\usepackage{cleveref}	
\usepackage{url}		
\usepackage{rotating}
\usepackage{stfloats}
\usepackage{longtable,lscape}
\usepackage{tabularx}

\newcommand{\kms}{\mbox{km\,s$^{-1}$}}
\def\kms{km${\rm s}^{-1}$}
\def\arcdeg{\hbox{$^\circ$}}
\def\arcmin{\hbox{$^\prime$}}
\def\arcsec{\hbox{$^{\prime\prime}$}}
\def\ergcms{erg\,cm$^{-2}$s$^{-1}$}
\def\micron{$\mu$m}
\def\ha{H$\alpha$}
\def\hb{H$\beta$}
\def\hg{H$\gamma$}
\def\hd{H$\delta$}
\def\NII{[N\,\textsc{ii}]}
\def\SII{[S\,\textsc{ii}]}
\def\SIII{[S\,\textsc{iii}]}
\def\SIV{[S\,\textsc{iv}]}

\def\OII{[O\,\textsc{ii}]}
\def\OIII{[O\,\textsc{iii}]}
\def\OIV{[O\,\textsc{iv}]}
\def\NeII{[Ne\,\textsc{ii}]}
\def\NeIII{[Ne\,\textsc{iii}]}

\def\ArIII{[Ar\,\textsc{iii}]}
\def\ArIV{[Ar\,\textsc{iv}]}

\def\HII{H\,\textsc{ii}}
\def\HeI{He\,\textsc{i}}
\def\HeII{He\,\textsc{ii}}
\def\CII{C\,\textsc{ii}}

\def\CIV{C\,\textsc{iv}}
\def\NIII{N\,\textsc{iii}}
\def\NIV{N\,\textsc{iv}}
\def\NV{N\,\textsc{v}}

\def\p0{\phantom{0}}

\def\lessim{\raise-.5ex\hbox{$\buildrel<\over{\scriptstyle\mathtt{\sim}}$}}
\def\grtsim{\raise-.5ex\hbox{$\buildrel>\over{\scriptstyle\mathtt{\sim}}$}}

\newcommand{\tick}{\ding{52}}

\newcommand{\notick}{\hspace{1pt}\ding{55}}

\title[A {[}WN{]} star in Abell 48]{The planetary nebula Abell 48 and its [WN] nucleus}
\author[D.J. Frew et al.]{David J. Frew$^{1,2}$\thanks{E-mail:david.frew@mq.edu.au}, I.S. Boji\v{c}i\'c$^{1,2,3}$,
Q.A. Parker$^{1,2,3}$, M. Stupar$^{1,2}$,  S. Wachter$^{4}$, \newauthor K. DePew$^{1,2}$, A. Danehkar$^{1,2}$, M.T. Fitzgerald$^{1,2}$ and D. Douchin$^{1,2}$\\
$^{1}$Department of Physics and Astronomy, Macquarie University, Sydney, NSW 2109, Australia\\
$^{2}$Research Centre in Astronomy, Astrophysics \& Astrophotonics, Macquarie University, Sydney, NSW 2109, Australia\\
$^{3}$Australian Astronomical Observatory, PO Box 915, North Ryde, NSW 1670, Australia\\     %
$^{4}$Spitzer Science Center, California Institute of Technology, MS 220-6, Pasadena, CA 91125, USA\\  
}

\begin{document}

\date{Accepted ; Received ; in original form }
\pagerange{\pageref{firstpage}--\pageref{lastpage}} \pubyear{}

\maketitle
\label{firstpage}

\begin{abstract}
We have conducted a detailed multi-wavelength study of the peculiar nebula Abell\,48 and its central star.  We classify the nucleus as a helium-rich, hydrogen-deficient star of type [WN4--5].  The evidence for either a massive WN or a low-mass [WN] interpretation is critically examined, and we firmly conclude that Abell\,48 is a planetary nebula (PN) around an evolved low-mass star, rather than a Population\,I ejecta nebula.  Importantly, the surrounding nebula has a morphology typical of PNe, and is not enriched in nitrogen, and thus not the `peeled atmosphere' of a massive star.  We estimate a distance of 1.6\,kpc and a reddening, $E(B-V)$ = 1.90\,mag, the latter value clearly showing the nebula lies on the near side of the Galactic bar, and cannot be a massive WN star.  The ionized mass ($\sim$0.3 $M_{\odot}$) and electron density (700 cm$^{-3}$) are typical of middle-aged PNe. The observed stellar spectrum was compared to a grid of models from the Potsdam Wolf-Rayet (PoWR) grid.  The best fit temperature is 71 kK, and the atmospheric composition is dominated by helium with an upper limit on the hydrogen abundance of 10 per cent.  Our results are in very good agreement with the recent study of Todt et al., who determined a hydrogen fraction of 10 per cent and an unusually large nitrogen fraction of $\sim$5 per cent.  This fraction is higher than any other low-mass H-deficient star, and is not readily explained by current post-AGB models.  We give a discussion of the implications of this discovery for the late-stage evolution of intermediate-mass stars. There is now tentative evidence for two distinct helium-dominated post-AGB lineages, separate to the helium and carbon dominated surface compositions produced by a late thermal pulse.  Further theoretical work is needed to explain these recent discoveries.
\end{abstract}

\begin{keywords}
planetary nebulae: general -- planetary nebulae: individual: Abell~48 -- stars: Wolf-Rayet -- stars: evolution 
\end{keywords}

\section{Introduction}

Classical Wolf-Rayet (WR) stars are massive hydrogen-deficient objects with powerful, fast winds and high mass-loss rates up to 10$^{-4}$ M$_{\odot}$ yr$^{-1}$ (Crowther 2007).  Their spectra are characterised by strong, broad  emission lines of  helium, nitrogen, carbon or oxygen.  A similar phenomenon is found in some ionizing stars of planetary nebulae (PNe), which descend from low mass progenitor stars and are unrelated in an evolutionary sense.  These central stars (CSPNe) are denoted as [WR] stars, utilising the square brackets introduced by van der Hucht et al. (1985) to avoid confusion with their massive analogues.  While massive WR stars are predominantly nitrogen or carbon enriched (WN or WC types respectively), with a few high temperature WO stars, almost all CSPNe catalogued to date belong to the [WC] $\rightarrow$ [WO] sequence (Tylenda et al. 1993; Crowther, De Marco \& Barlow 1998; Acker \& Neiner  2003), with one or two exceptions (e.g. Miszalski et al. 2012b, hereafter MC12).

\begin{figure*}
\begin{center}
\includegraphics[width=16.cm]{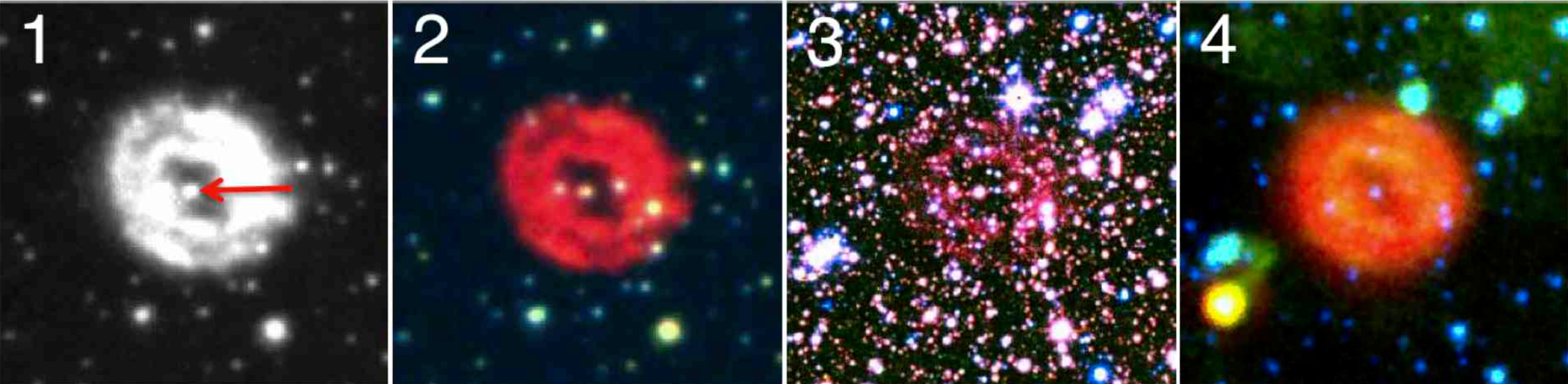}
\caption{Montage of images for Abell~48, all to the same scale and orientation (90\arcsec\,$\times$\,90\arcsec\ with NE at top left), and with the colour palette set so that RGB channels map longer to shorter wavelengths respectively.  1. SHS \ha\ image with the CSPN marked -- note the elliptical, double-ring structure of the nebula; 2. SSS/SHS composite B$_{J}$/R/H$\alpha$ colour image; 3. UKIDSS  $JHK_{s}$ composite  image; 4. IRAC/MIPS composite 4.5/8.0/24\,\micron\ image.  A colour version of this figure is available in the online journal.  }
\label{fig:montage}
\end{center}
\end{figure*}

In the course of a spectroscopic survey of detectable CSPNe (DePew et al. 2011), we were struck by the unusual nature of the nucleus of Abell~48 (PN~G029.0+00.4), discovered and catalogued as a PN by Abell (1955, 1966).  Abell~48 is a reddened, low-surface brightness PN with a morphology resembling a thick torus or cylinder, possibly viewed nearly pole-on (Figure\,\ref{fig:montage}).    
Abell (1966) classified it as a double-ring nebula\footnote{Curiously, Helfand et al. (2006) classified it as a candidate supernova remnant.} and gave an optical diameter of 40\arcsec.  It has not been well studied optically due to its relative faintness.
DePew et al. (2011) already concluded that Abell~48 was a PN and gave a preliminary classification of the CSPN as [WN] or [WN/C], but since the important CIV $\lambda$5806 doublet was unobserved in that work, a more precise spectral type could not be determined.  Independently,  Wachter et al. (2010) classified the central source as a  WN6 star, in contrast to the object's previous long association as a PN  (Perek \& Kohoutek 1967; Zuckerman \& Aller 1986; Acker et al. 1992; Kohoutek 2001).   This uncertainty led us to more closely investigate this interesting object and tie down its identification.

In this paper we present a detailed multi-wavelength study of Abell 48 and its CSPN, expanding on the preliminary analysis reported in Boji\v{c}i\'c et al. (2013).  As this paper was completed, an independent study of this object was published by Todt et al. (2013, hereafter TK13), coming to essentially the same conclusions as us.   We will refer to their results in the context of our own paper where necessary.  We have confirmed Abell 48's CSPN  as a hydrogen-poor [WN] star, as first suggested by DePew et al. (2011).   We also explore the nature and characteristics of the [WN] class based on a comparison of Abell~48 with the other  [WN] and [WN/C] stars, their possible progenitors, and their likely progeny (Todt et al. 2010a,b, 2012; Werner 2012; MC12; TK13).   This paper is arranged as follows: we describe our observations of Abell~48 in \S~2, undertake a detailed analysis of the nebula and its CSPN in \S~3 and \S~4, and discuss the evidence for its nature in \S~5.  In \S~6 we review the [WN] or [WN/C] classes, discuss the possible evolutionary pathways of these groups in \S~7, before giving our conclusions and suggestions for future work in \S~8.

\section{Observations}

\subsection{Spectroscopic Observations}

Optical spectra of  Abell 48 and its CSPN were obtained on the ANU 2.3-metre telescope at Siding Spring Observatory using the Wide Field Spectrograph (WiFeS; Dopita et al. 2007, 2010), on 1 July 2009.  WiFeS has a double-beam configuration, with spectral resolutions of 3000 or 7000, and a 38\arcsec\ $\times$ 25\arcsec\  field of view with a spatial resolution of 1\arcsec.  Since the angular size of Abell 48 is larger than this, the instrument was positioned on the northern section of the nebula (including the CSPN) to make sure that some suitable background sky was available for subtraction.  This was an early run during WiFeS commissioning time so the more effective nod-and-shuffle mode was not used.   Low resolution gratings (B3000 and R3000) were employed and the exposure time on target was 1500s.  Thin cloud was present during the observation.

\begin{table*}
{\footnotesize	
\begin{center}
\caption{Spectroscopic and photometric observations of Abell 48 utilised in this study.}
\label{table:log}
\begin{tabular}{lcccccc}
\hline
Date    			& Telescope 		 		& Instrument   			&     $\lambda$ Range  			& $R\,= \lambda/\Delta\lambda$ 	&  Exp. time       	  	 			\\
\hline
2008 Sep 4        	&  Hale 200-in   			& Dual Spectrograph		& 	3400--5600, 5850--8300		&     1000		 		&      900			  				\\
2009 Jul 1         	& ANU 2.3-m	 			& WiFeS					&   	3400--5900, 5300--9600 		&	2900		  		 &	 1500				  		\\
2010 Apr 22      	&  ANU 2.3-m	 			& WiFeS					&   	4180--5580, 5300--7060 		&	6900		 		 &	 1200				 	 	\\
2012 Aug 23      	&  ANU 2.3-m	 			& WiFeS					&   	3400--5900, 5300--9600 		&	2900		 		 &	 1200				 		\\
\hline
Date    			& Telescope 		 		& Instrument   			&     Filter		 				&  	...				 	&  Exp. time       	  	 			\\
\hline
2012  Feb 24	 &  2.0-m FTN	 				& Merope Camera		&   	$B	$			 			&	...		 		 	&  	  900						\\
2012   Mar 14	&  2.0-m FTN	 				& Merope Camera		&   	$I_c$			 			&	...		 		 	&  	  10							\\
2012   May 8		 &  2.0-m FTN	 			& Merope Camera		&   	$BV$			 			&	...		 		 	&  	200, 150						\\
2012  May 18    	&  ANU 2.3-m	 			& Imager				&   	$UBVI_c$			 		&	...		 		 	& 300, 2$\times$300, 2$\times$180, 2$\times$120 \\
\hline
\end{tabular}
\end{center}
}
\end{table*}

Higher resolution gratings (B7000 and R7000) were used on a second run in April 2010.  This observation was aimed at getting kinematic data on the nebula, so the IFU was positioned to cover both the CSPN and centre of the nebula.  A further exposure of the southern section of the nebula was taken using WiFeS at low resolution in August 2012.  The night was photometric and consequently the S/N ratio is best on this spectrum.
For the WiFeS data, the frames were bias subtracted and flux- and wavelength-calibrated using the WiFeS data reduction pipeline (Dopita et al. 2010) in conjunction with standard {\sc IRAF} routines.  The flux calibration was performed using the spectrophotometric standard  LTT\,3864.  

We also utilised the spectrum of the CSPN first presented by Wachter et al. (2010), obtained with the 5.0-m Hale telescope at Palomar Observatory on 4 September 2008 using the Double Spectrograph. 
This spectrograph also has a two-arm configuration, utilising a dichroic to split the light in to two channels, observed simultaneously.  The 316 line mm$^{-1}$ grating in first order, dichroic D55 and a slit width of 1\arcsec\ were used.  The resulting spectra cover 4000--5600 \AA\ and 5800--8300 \AA\ with a dispersion of 2.0 and 2.4 \AA\ pixel$^{-1}$ on the blue and red sides, respectively, with a spectral resolution of 5--7 \AA.  ~FeAr (blue) and HeNeAr (red) arcs were used for wavelength calibration.  Bias and flat-field corrections to the raw spectrum images were performed using standard {\tt IRAF} routines.  A log of the observations is presented in Table~\ref{table:log}.

\subsection{Photometric Observations}

New optical photometry of the CSPN was also obtained with the Australian National University's 2.3-m reflector at Siding Spring Observatory (SSO) in May 2012,  and the 2.0-m Faulkes Telescope North (FTN) at Haleakala Observatory, Maui, in queue mode on three different nights in 2012.  A log of the observations is presented in Table~\ref{table:log}.
The SSO observations used  the standard imager at the f/18 Nasmyth focus.  This used a 2048$\times$2048 pixel thinned E2V CCD with a pixel size is 13.5\,$\mu$m, giving a plate scale of 0.34\arcsec/pixel  across a 6.6\arcmin\  circular field of view. Standard $UBVI_c$ filters were used.  The seeing was modest ($\sim$1.5\arcsec) and the night was photometric.  Standard extinction coefficients were applied, and a range of standard stars from Landolt (2009) were measured to determine the colour equations used to convert from instrumental magnitudes.  
For the FTN observations, the Merope camera at the f/10 focus was used with $BVI_C$ Johnson-Cousins filters.  The  camera utilised an E2V CCD with 2048 $\times$ 2048 pixels, in 2\,$\times2$ binning mode, giving images 4.7\arcmin\ on a side with a plate scale of 0.28\arcsec/pixel.    The averaged observed magnitudes are presented in \S~\ref{sec:photometry}.

\subsection{Archival Data}

To supplement our spectroscopic and photometric data, we made an extensive search for archival multi-wavelength data of the nebula and central star using the Aladin Sky Atlas\footnote{Accessible from the Centre de Donn\'es Astronomiques  (CDS).}, and the SkyView Virtual Observatory\footnote{\url{http://skyview.gsfc.nasa.gov/}}.  Narrowband \ha\ (+ \NII) images were obtained from the SuperCOSMOS H$\alpha$ Survey (SHS; Parker et al. 2005; Frew et al. 2013b) and the Southern H-Alpha Sky Survey Atlas (SHASSA) (Gaustad et al. 2001).   Broadband images were also obtained from the SuperCOSMOS Sky Survey (SSS; Hambly et al. 2001), the 2MASS (Skrutskie et al. 2006) and UKIDSS (Lawrence et al. 2007) surveys in the near-IR, and the GLIMPSE (Benjamin et al. 2003; Churchwell et al. 2009), MIPSGAL (Carey et al. 2009), and WISE (Wright et al. 2010) mid-IR surveys.  

Continuum fluxes for the nebula (and the CSPN separately if available), were retrieved from the VizieR service or measured from the original images as part of this work.  We also downloaded 20\,cm and 90\,cm high-resolution MAGPIS images (Helfand et al. 2006) to supplement the NVSS radio data (Condon \& Kaplan 1998).  Figure~\ref{fig:montage} shows a multiwavelength montage of merged-colour images for Abell~48, all to the same scale.  We adopt the best position for the CSPN from the 2MASS catalogue, $\alpha$ = 18$^{\rm h}$\,42$^{\rm m}$\,46.92$^{\rm s}$ , $\delta$  = $-03$\arcdeg\ 13\arcmin\ 17.3\arcsec\ (J2000).    

\section{The Planetary Nebula}

\subsection{Nebular Morphology}
\begin{figure}
\begin{center}
\includegraphics[height = 5.1cm]{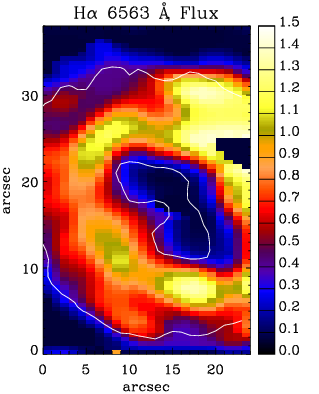}
\includegraphics[height = 5.1cm]{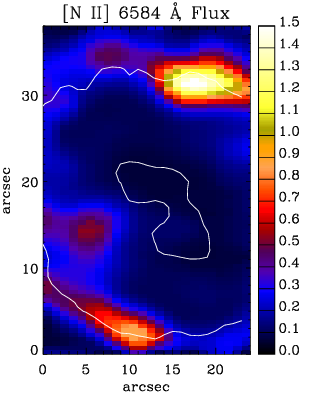}
\caption{Surface brightness maps of Abell~48 in the \ha\ and \NII$\lambda$6584 lines, extracted from the WiFeS observation of Apr 2010.  Note the pair of opposing knots which are prominent in the \NII\ $\lambda$6584 line. North-east is at top left in each panel and the flux units are in $10^{-15}$\,\ergcms\ per spaxel.  A colour version of this figure is available in the online journal.  }
\label{fig:A48_knots}
\end{center}
\end{figure}

Abell\,48 has an apparent `double-ring' morphology with a lozenge-shaped interior cavity (see Figure~\ref{fig:montage}), which might be the projection of a thick torus or cylinder viewed nearly pole-on, as has been kinematically demonstrated for other morphologically similar nebulae (e.g. O'Dell et al. 2013).  The measured dimensions from the SHS are 44\arcsec $\times$ 39\arcsec.      
There is also a pair of point symmetric knots  which are prominent in the \NII\ $\lambda$6584 line (see Figure~\ref{fig:A48_knots}), and which are often seen in PNe.  This fact, the PN-like morphology seen in \ha\ images, and the emission measure of the ionized nebula all indicate that it is material recently ejected by the CSPN, and not simply windswept interstellar material (see the morphological atlas of WR shells by Gruendl et al. 2000).  This fact has important implications for the origin of the nebula, as we elucidate below. 

Soker (1997) hinted that the CSPN of Abell~48 may be a binary system, based on nebular morphology (see later).  If this turns out to be true, a binary evolution channel may be implicated to explain the unusual surface abundances of the CSPN, as has been suggested for the LMC PN, N\,66 (Hamann et al. 2003) and for the peculiar CSPN of the Eskimo nebula, NGC~2392 (e.g. M\'endez et al. 2012).   The nebular emission-line profiles (see section~\ref{sec:exp_vel}, below) also appear to rule out simple spherical symmetry (cf. L\'opez et al. 2012), so we tentatively interpret the double shell morphology as due to the geometric orientation of two opposing lobes.  A full morpho-kinematic analysis of the nebula, based on our WiFeS imaging data, will be the subject of a future paper, but a preliminary analysis indicates this is the likely scenario at play.

\begin{figure}
\begin{center}
\includegraphics[width =7.25cm]{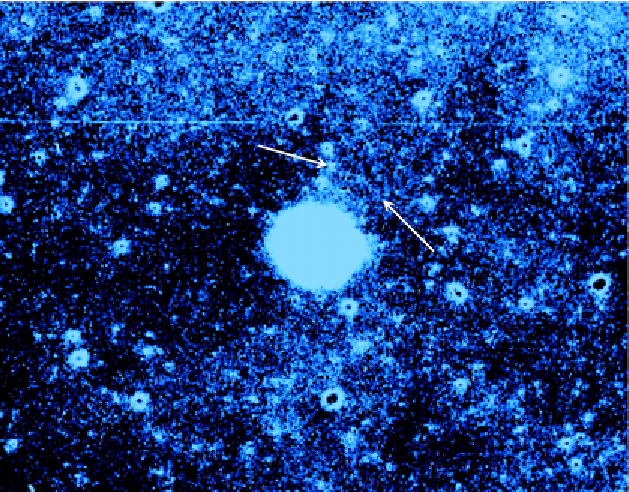}
\caption{A deep continuum-divided (H$\alpha$) SHS image of Abell 48 showing two short arcs, the outer one arrowed, located 35\arcsec\ and 45\arcsec\ north-west of the CSPN.  The image subtends 300\arcsec\,$\times$\,240\arcsec\ with NE at top left.}
\label{fig:A48_halo}
\end{center}
\end{figure}

To the northwest of the PN there appear to be two faint closely spaced arcs about 30\arcsec\ in extent, with the outermost being 45\arcsec\  from the CSPN (see Figure~\ref{fig:A48_halo}).  These arc-like structures were first noted by DePew (2011) and there also appears to be evidence of a very faint elliptical halo on deep continuum-subtracted SHS \ha\ images, discovered as part of the survey of Frew, Boji\v{c}i\'c \& Parker (2012)\footnote{The SHS is an excellent search medium for faint structures surrounding PNe and symbiotic stars (see also Miszalski et al. 2012a).}.  The overall dimensions are about 280\arcsec\ $\times$ 210\arcsec\ in extent, but deep CCD images are needed to confirm them.  Together, these features may provide evidence for earlier AGB mass-loss events.   
Even though Abell~48 is close to the Galactic mid-plane, where the interstellar medium (ISM) density is highest (Spitzer 1978), the PN has little signature of an ISM interaction  (Soker, Borkowski \& Sarazin 1991; Pierce et al. 2004; Wareing et al. 2006; Wareing 2010; Frew et al. 2011; Ali et al. 2012).  Therefore the peculiar velocity of the CSPN is likely to be low ($<$40\,\kms).   In contrast, TK13 claim Abell~48 is a runaway star, based on the CSPN's proper motion of 19.5 mas\,yr$^{-1}$ in p.a. 222\arcdeg\ from the PPMXL catalogue (Roeser, Demleitner \& Schilbach 2010).  However, this  differs from the USNO-B1.0 catalogue value (Monet et al. 2003) of 25.3 mas\,yr$^{-1}$ in p.a. 342\arcdeg.  Owing to the faintness of the star, these catalogue values may be spurious.

\subsection{Integrated Nebular Fluxes}\label{nebular_fluxes}   
We have four independent estimates of the integrated \ha\ flux.  Frew et al. (2013a) derived log$F$(\ha) = $-$11.52 $\pm$ 0.09 (in cgs units) from aperture photometry on SHASSA images, while Frew et al. (2013b) estimated log$F$(\ha) = $-$11.56 $\pm$ 0.15 from newly-calibrated SuperCOSMOS \ha\ images, in good agreement.  
We also used our WiFeS data, scaling up the fluxes to account for the portion of the nebula that was not observed.  We did this for two separate runs to obtain log$F$(\ha) = $-11.59$ and $-11.62$ respectively.    The weighted mean flux  is log\,$F$(\ha) = $-11.58 \pm 0.06$ \ergcms, adopted hereafter.

\begin{table}                
{\footnotesize	
\begin{center}
\caption{Summary of continuum flux measurements for Abell 48.}
\label{table:continuum}
\label{phot}
\begin{tabular}{lllc}
\hline
Wavelength & Flux  & Survey & Reference\\
($\mu$m)  & (mJy)&   &  \\
\hline								
3.6      &        67 $\pm$ 7  		         & IRAC &       C11     \\
4.5     &      109 $\pm$ 10		         & IRAC &       C11      \\ 
5.8     &      32 $\pm$ 5 		          & IRAC &       C11     \\ 
8.0     &     244 $\pm$  24       	          & IRAC &       C11      \\ 
8.3     &	    375 	$\pm$ 15$^{a}$		    &	MSX     &       EP03  \\  
11.6     &	 	1195 $\pm$ 120:	&	WISE  &  This work   \\
14.7      &		2053 $\pm$ 125	 &	MSX   & EP03   \\  
18         &       3341  $\pm$ 42  		&	AKARI  &I10   \\  
21.3     &		1735  $\pm$ 110	&	MSX &  EP03    \\  
22.1     & 	       1600 $\pm$ 200	&	WISE &This work    \\
24         &        2065 $\pm$ 100           &   MIPSGAL  &  M10, PM11      \\ 
25          &	    2530 $\pm$ 250 		&	IRAS &  IPAC \\  
60           &		$<$2070 			&	IRAS  &  IPAC  \\  
65           &    29020  $\pm$ 570$^a$		&	AKARI & I10  \\ 
70          &            11000    $\pm$ 2000         &   MIPSGAL  &  This work      \\ 
90           &		30990 $\pm$ 2200$^{a}$		&	AKARI & I10  \\ 
\hline
6 cm          &      $>$70                          &     MAGPIS      &  This work   \\   
6 cm          &      $<$400                          &     NVSS     &  This work  \\  
11 cm       &     $\ge$220                           &      VLA   & PB03, this work    \\     
20 cm       &         159   $\pm$ 15             &       NVSS  & C98  \\   
20 cm       &             200 $\pm$ 20            &       MAGPIS  & This work   \\   
90 cm       &             60  $\pm$ 10            &	MAGPIS	 & This work   \\	
\hline
\end{tabular}
\end{center}
\begin{flushleft}
References:  C11 --  Cohen et al. (2011); C98 -- Condon et al. (1998); EP03 -- Egan et al. (2003);  I10 -- Ishihara et al. (2010);  IPAC -- IPAC (1986); M10 -- Mizuno et al. (2010);  PB03 -- Paladini et al. (2003); PM11 -- Phillips \& Marquez-Lugo (2011); 
$^{a}$  Fluxes likely confused with surrounding emission.
\end{flushleft}
}
\end{table}

In the radio-continuum domain we found two catalogued but discrepant values for the 1.4\,GHz (20 cm) integrated flux density of Abell 48;  $S_{1.4}$ = 159 $\pm$ 15~mJy from the NRAO VLA Sky Survey (NVSS; Condon et al. 1998; Condon \& Kaplan 1998) and $S_{1.4}$ $\approx657$~mJy, catalogued in the Multi-Array Galactic Plane Imaging Survey (MAGPIS; Helfand et al. 2006). The discrepancy is too large to be attributed to confusion by the complex background,  as noted by Condon \& Kaplan (1998). Therefore, we re-measured the flux density from the MAGPIS and NVSS total intensity images\footnote{The images were retrieved from the NVSS and MAGPIS postage stamp servers: \url{http://www.cv.nrao.edu/nvss/} and \url{http://third.ucllnl.org/gps/}} using the {\tt KARMA} data analysis package (Gooch 1996).  While our estimated flux density from the NVSS image agrees with the catalogued value within the uncertainties, we measured a flux of $\sim$200\,mJy from the MAGPIS image, at odds with the catalogued value, but closer to the NVSS flux.  In fact some published MAGPIS fluxes need to be revised (D. Helfand, 2011, private communication), and the catalogued value of 657\,mJy is erroneous in this case. 
Abell~48 is not detected in the high-resolution Coordinated Radio and Infrared Survey for High-mass Star Formation (CORNISH) Source Catalog (Purcell et al. 2013), recently undertaken at 5\,GHz with the Very Large Array (VLA). Abell~48 is just below the survey sensitivity threshold and is likely resolved out by the $uv$-coverage of the VLA B-configuration (C. Purcell, 2013, priv. comm.).  
The resolution of the MAGPIS image allows us to filter out the features close to Abell 48 and so avoid the confusion problem which affects the older NVSS data.  Hence, we use our new measurement for further analysis. In addition, we measured the flux density at 0.325\,GHz (90~cm) from an intensity image taken as a part of the MAGPIS survey (for more details see Helfand et al. 2006). Both flux values are tabulated in Table~\ref{table:continuum}, along with several additional integrated flux measurements of the nebula taken from the literature.  It is clearly seen from these fluxes that the emission from the nebula is thermal in nature, typical of photoionized gas.
\subsubsection{Infrared Fluxes}\label{sec:MIR}

These are several flux determinations at infrared (IR) wavelengths.  The nebula is faint in 2MASS and UKIDSS images.  In the $K_s$ band, which has the strongest detection, we attribute the flux to molecular hydrogen (H$_2$) at 2.12\,$\mu$m, as well as the Br$\gamma$ 2.17\,$\mu$m hydrogen line (Kimeswenger et al. 1998).  The GLIMPSE data resolves the nebula well, and it appears brightest in the IRAC2 and IRAC4 bands.  On this basis, the Br$\alpha$ 4.05 $\mu$m line is probably the strongest feature contributing to the IRAC2 flux, and strong polycyclic aromatic hydrocarbon (PAH) bands (Allamandola, Tielens \& Barker 1989) likely contribute to the IRAC4 flux, particularly the bands at 7.7 and 8.6\,$\mu$m, with some emission from Pf\,$\alpha$ 7.46$\mu$m and the \ArIII\ 8.99$\mu$m fine-structure line.   We ran a preliminary photoionization model to test these predictions using the three-dimensional photoionization code {\tt MOCASSIN} (Ercolano et al. 2003, 2005).  This predicts an \ArIII\  8.99$\mu$m line flux of around 20\% of \hb.  A full analysis will be published separately.

Abell~48 is prominent in the WISE3 (11.6\,\micron) and WISE4 (22\,\micron) bands, but only barely seen in WISE2.  Features that may contribute to the WISE3 band are PAH emission, particularly at 11.3\,$\mu$m, thermally-emitting dust, and the \SIV\ 10.5\,$\mu$m line which our photoionization model suggests is comparable in strength to \hb.   Other predicted contributors, in order of decreasing strength, are the \ArIII\ 8.99$\mu$m,  \NeIII\ 15.5$\mu$m, and \NeII\ 12.8$\mu$m lines.  Similarly, it is a moderately strong source in the MIPS 24\,\micron\ band (Mizuno et al. 2010; Wachter et al. 2010), considering that the \OIV\ fine-structure line is expected to be weak or absent in a nebula of this excitation class.  In fact our optical spectra (Figure~\ref{A48_nebula_spectrum}) show  no  emission lines with an ionization potential (I.P.) exceeding 41.0\,eV, based on the weak detection of the \NeIII\  $\lambda$3869 line. 
This suggests that warm dust is the main contributor to the 22 and 24\,\micron\ bands, with a likely minor contribution from the \ArIII\ 21.81$\mu$m line.  The spectral energy distribution (SED) of the nebula, not shown here, is dominated by this dust component.  From a blackbody fit to the IR fluxes given in Table~\ref{table:continuum} we estimate a dust temperature of 150 K,  quite typical of PNe.  Mizuno et al. (2010) also found that the nebular 8\,\micron\ emission was co-spatial with the 24\,\micron\ emission.  Similarly, we find the angular size to be similar at all wavelengths indicating that the ionized, molecular, and dust components are spatially mixed.  The  IRAC/MIPS false-colour  image\footnote{MIR false-colour imaging is also a powerful diagnostic indicator for classification purposes (e.g. Cohen et al. 2007, 2011; Parker et al. 2012).} (Figure~\ref{fig:montage}) illustrates this.  A  summary of all nebular continuum fluxes  is given in Table~\ref{table:continuum}.

\subsection{Nebular Spectroscopy}\label{sec:nebular_spectroscopy}
\begin{figure*}
\begin{center}
\includegraphics[width=13.75cm]{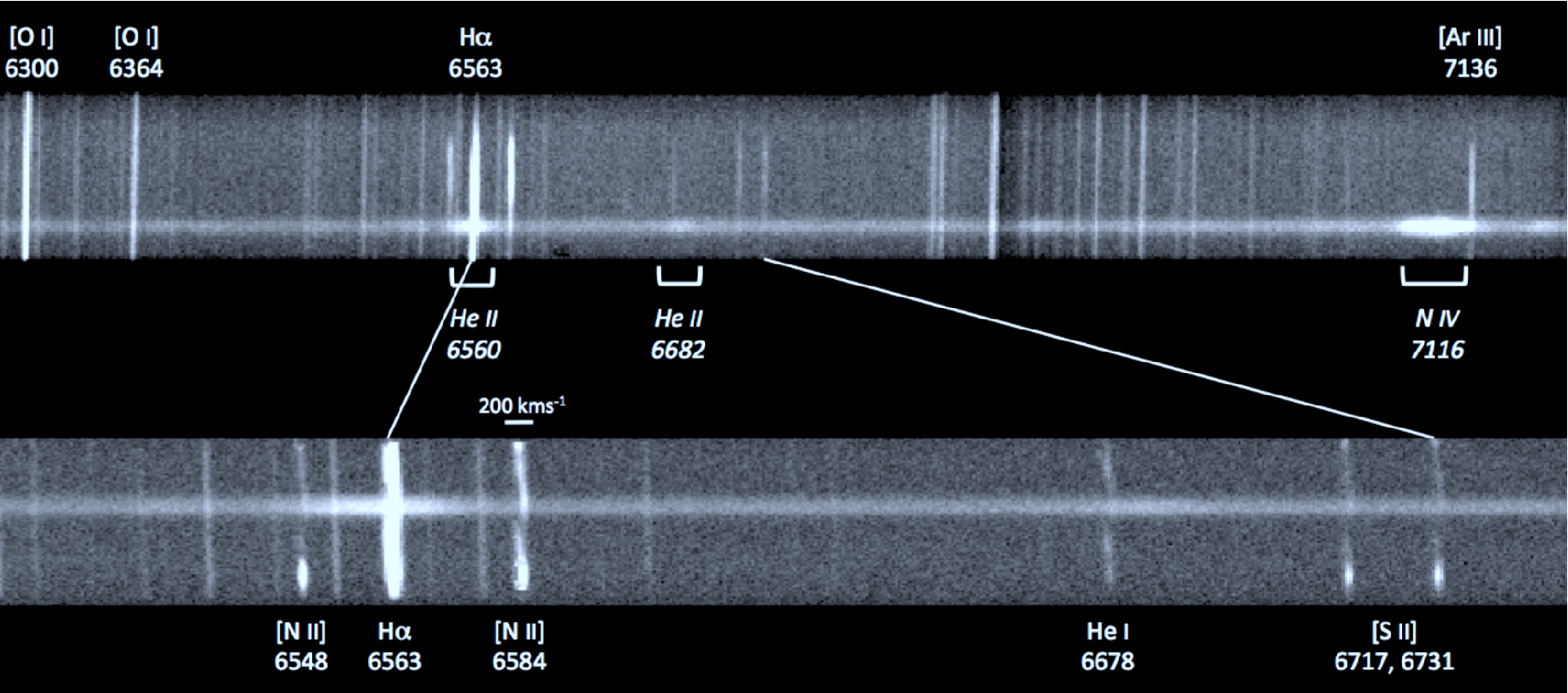}
\caption{Individual image slices of our low-resolution (R = 3000; top panel) and medium-resolution (R = 7000; bottom panel) WiFeS red spectra showing the broad features from the CSPN, and line-splitting of the nebular lines due to expansion.}
\label{2Dspec}
\end{center}
\end{figure*}

\begin{figure*}
\begin{center}
\includegraphics[height=6.2cm]{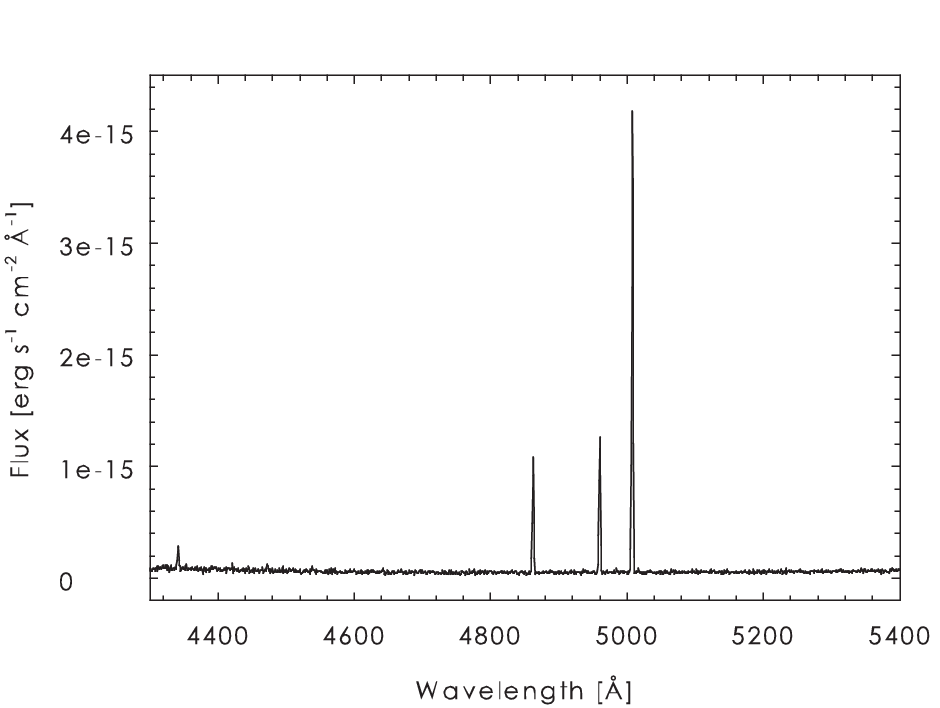}
\includegraphics[height=6.2cm]{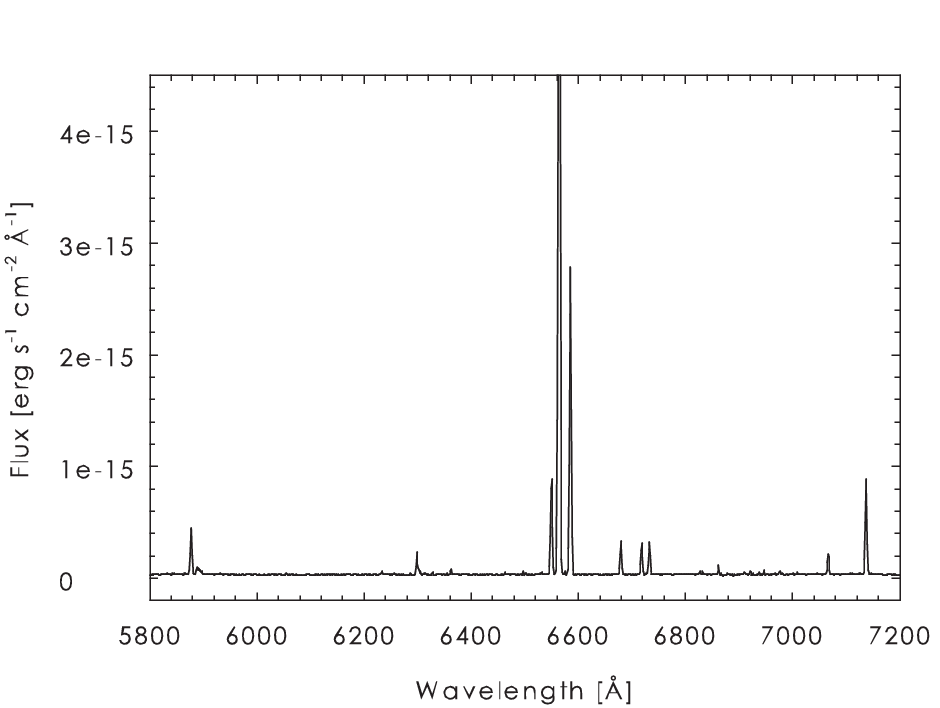}
\caption{Extracts of the integrated blue (left) and red spectra of Abell 48, obtained from the high S/N WiFeS observation from 2012. Note that the \ha\ line is truncated in the right panel. The observed \NII/\ha\ ratio is 0.28.}
\label{A48_nebula_spectrum}
\end{center}
\end{figure*}

Our three WiFeS data cubes were used to investigate the nebular conditions. Figure~\ref{2Dspec} presents individual slices of the low-resolution (R = 3000) and medium-resolution (R = 7000) WiFeS red-arm spectra respectively. The broad features from the CSPN, the respective nebular and night sky line strengths, and line-splitting of the nebular lines due to expansion can be seen.
The 1-D low-resolution spectrum from 2012, binned from the data cube to increase the signal to noise (S/N) ratio, is illustrated in Figure~\ref{A48_nebula_spectrum}.  It shows the presence of a bright nebular \ha\ line and relatively weak \NII\ $\lambda \lambda$\,6548,84 lines in the red (\NII/\ha\ = 0.28), a steep Balmer decrement due to the relatively high extinction, as well as fairly weak {[\rm O \sc iii]} lines in the blue relative to \hb.  There is no indication of a nebular \HeII\ $\lambda$4686 line (to a level of 1\% percent of \hb).

\begin{table}
{\footnotesize	
\begin{center}
\caption{List of emission line fluxes for Abell 48, adopted  from the high S/N WiFeS IFU spectrum from 2012.}
\label{table:spectroscopy}
\begin{tabular}{lcrrr}
\hline
Line	&$\lambda$&$f$($\lambda$)& ~$F$($\lambda$)~	&~$I$($\lambda$)~	\\
\hline									
\OII	 	&~~3727~~	&	0.256	&	13:		&     64:	\\
\NeIII 	&	3869	&	0.230	&	6.8:		&     28:	\\
\hd\		&	4101	&	0.182	&	6.6:		&	20:	\\
\hg\		&	4340	&	0.127	&	19.9	       &	43.5 	\\
\OIII 	&	4363	&	0.121	&	$<$1.0	&     $<$2.1	\\
\HeI		&	4471	&	0.095	&	5.0: 		&     9.0:	\\
\HeII		&	4686	&	0.043	&	$<$0.8 	&     $<$1.0	\\
\hb\		&	4861	&	0.000	&	100		&	100	     \\
\OIII 	&	4959	&	-0.024	&	119		&	103	\\
\OIII 	&	5007	&	-0.036	&	402		&	322	\\
\NII	 	&	5755	&	-0.195	&	1.3:		&	0.40: \\
\HeI		&	5876	&	-0.215	&	67.6		&	17.9 	\\
\SIII		&	6312	&	-0.283	&	2.1:		&	0.37:	\\
\NII	 	&	6548	&	-0.318	&	138		&	19.4	\\
\ha\		&	6563	&	-0.320	&	2056	&	286	\\
\NII	 	&	6583	&	-0.323	&	429		&	58.4	\\
\HeI		&	6678	&	-0.336	&	43.7		&	5.5	\\
\SII		&	6716	&	-0.342	&	42.3		&	5.1	\\
\SII		&	6731	&	-0.344	&	44.1		&	5.3	\\
\HeI		&	7065	&	-0.387	&	28.0		&	2.6	\\
\ArIII	&	7136	&	-0.396	&	129		&	11.2	\\
\CII\		&	7236	&	-0.409	&	12.9:	&	1.0:	\\
\HeI		&	7281	&	-0.414	&	12.1		&	0.94	\\
\OII		&	7320	&	-0.419	&	10.3:	&	0.78:	\\
\OII		&	7330	&	-0.420	&	7.2:		&	0.54:	\\
\ArIII	&	7751	&	-0.467	&	33.8		&	1.9	\\
{[Cl\,\textsc{iv}]}& 8046 & -0.497	&	2.0:		&	0.1: \\
P16		&	8502	&	-0.540    &	25:		&	0.9:	\\
P15		&	8545	&	-0.542	&	16:		&	0.6:	\\
P14		&	8599	&	-0.547	&	13:		&	0.4:	\\
P13		&	8665	&	-0.553	&	29:		&	0.9:	\\
P12		&	8750	&	-0.560	&	29:		&	0.9:	\\
P11		&	8863	&	-0.569	&	35:		&	1.0:	\\
P10		&	9015	&	-0.581	&	20:		&	0.6:	\\
\SIII		&	9069	&	-0.585	&	787		&	21.3	\\
P$\zeta$	&	9229	&	-0.597	&	94.6		&	2.4	\\
\hline		
\end{tabular}
\end{center}
}
\end{table}

In Table~\ref{table:spectroscopy}, we summarise the observed  and reddening-corrected line fluxes.  The other WiFeS spectra were used to confirm that the faintest nebular lines were real.  The spectrum presented by TK13 is similar to our own, the main difference being that TK13 estimated a somewhat higher extinction.  Abell~48 has a low excitation class, EC = 1.45 (Dopita \& Meatheringham 1990; Reid \& Parker 2010), so we conclude that the weak $\lambda$7236 feature that we also detected is more likely to be a \CII\ recombination line rather than the  \ArIV\ line on the basis of our photoionization model.
  
\subsubsection{Reddening}\label{sec:reddening}
A summary of the various reddening determinations is presented in Table~\ref{table:reddening}.  We derived the logarithmic extinction at \hb, and hence $E(B-V)$, using the nebular Balmer decrement from our three independent WiFeS spectra, where the intrinsic line strengths from Hummer \& Storey (1987) and the reddening law of Howarth (1983) were used.  We independently determined the extinction by comparing the integrated 1.4\,GHz and \ha\ fluxes from \S~\ref{nebular_fluxes} (see, e.g. Boji\v{c}i\'c et al. 2011b).  We also utilised the P$\zeta$/\ha\ decrement, again adopting the line strengths from Hummer \& Storey (1987).    

Following Crowther et al. (2006a), interstellar $K_S$-band extinctions were also estimated from the observed $J-K_s$ and $H-K_s$ colours of the ionizing star, derived from Table~\ref{table:CS_phot}.  To estimate the intrinsic colours,  we take the average of the values for the weak-lined WN\,3--4 and WN\,5--6 subgroups from Crowther et al. (2006a), i.e. $(J-K_s)_0$ = +0.04 and $(H-K_s)_0$ = +0.07.  Using the recipe of Crowther et al. (2006a), we calculate $K_S$-band extinctions of $A_{K_s}^{J-K_s}$ = 0.78 and $A_{K_s}^{H-K_s}$ = 0.84, respectively.  Adopting the mean value, $A_{K_s}$ = 0.81, and using the relation between $A_{K_s}$ and $E(B-V)$ from Indebetouw et al. (2005), we derive $E(B-V)$ = 2.3 $\pm$ 0.3, in fair agreement with the other determinations.  An independent, albeit lower accuracy estimate is derived from the equivalent width (2.6\AA) of the $\lambda$6284 diffuse interstellar band (DIB) in the red (the only one measurable), and applying the relation of Friedman et al. (2011).   Finally, we re-evaluated the estimate of $A_{V}$ = 5.7 mag from Wachter et al. (2010) by fitting the overall SED of the star, excluding the uncertain IRAC [8.0]-mag value,  to determine $E(B-V)$ = 1.90 $\pm$ 0.10 (see \S~\ref{sec:stellar_SED}).  We adopt an averaged reddening, $E(B-V)$ = 1.90 $\pm$ 0.10 ($c_{\rm H\beta}$ = 2.75 $\pm$ 0.15) hereafter.  This value can also be compared with the estimate of $E(B-V)$ = 2.1 ($c_{\rm H\beta}$ = 3.0) from DePew et al. (2011), based on the nebular Balmer decrement from a shallow MSSSO 2.3-m spectrum, while TK13 determined $E(B-V)$ = 2.10 from a SED fit, and $E(B-V)$ = 2.15 from the Balmer decrement method.

\begin{table}
\caption{A summary of the various reddening determinations for Abell 48 which are all in good agreement.}
\begin{center}
\begin{tabular}{l c }
\hline
Method 							& ~~$E(B-V)$  \\ [0.5ex]
\hline
Stellar SED 						& ~~1.90 $\pm$ 0.10 \\
Near-IR stellar colours~~~~~~~ 	& ~~2.3 $\pm$ 0.3 \\
Radio/\ha\ flux 					&  ~~1.77 $\pm$ 0.15   \\
\ha/\hb\ decrement$^{1}$ 			& ~~2.2 $\pm$ 0.2 \\
\ha/\hb\ decrement$^{2}$ 			& ~~1.92 $\pm$ 0.1 \\
\ha/\hb\ decrement$^{3}$ 			& ~~1.85 $\pm$ 0.1 \\
P$\zeta$/\ha\  decrement$^{1}$ 	& ~~2.2 $\pm$ 0.2\\
P$\zeta$/\ha\  decrement$^{3}$ 	& ~~1.8 $\pm$ 0.2\\
$\lambda$6284 DIB  				&  ~~1.9 $\pm$ 0.5 \\
\hline
Adopted reddening				&  ~~1.90 $\pm$ 0.15 \\
\hline
\end{tabular}
\end{center}
\begin{flushleft}
\footnotesize{$^{1}$From WiFeS spectrum 1; $^{2}$WiFeS spectrum 2; $^{3}$WiFeS spectrum 3.}
\end{flushleft}
\label{table:reddening}
\end{table}

\subsubsection{Plasma Diagnostics}\label{sec:plasma}
We estimated the nebular density of Abell~48 to be $n_e$ = 700~cm$^{-3}$ from the observed ratio of the \SII\ doublet.  The \NII\ $\lambda$5755 line is barely detected giving an uncertain value of the electron temperature, while the far red \SIII\ lines were used to estimate $T_{\rm e}$ = 7000~K, and another low quality estimate comes from the ratio of the auroral to nebular \OII\ lines.  
Two further estimates come from the relative strengths of the \HeI\ lines (Zhang et al. 2005).   Despite the larger uncertainties of some of the fainter diagnostic lines, the independent estimates of the electron temperature are in good agreement, and we adopt an average value of 7500\,K for the abundance analysis (\S~\ref{sec:A48_abundances}).   A summary is presented in Table~\ref{table:plasma}.

\begin{table}
{\footnotesize	
\begin{center}
\caption{Summary of plasma diagnostics for Abell~48}
\label{table:plasma}
\begin{tabular}{lll}
\hline									
Diagnostic												&	~~~~Value				&   	~~~Result							\\
\hline		
\SII\	$\lambda$6717/$\lambda$6731   						&    ~~~~~~~0.96  			&    ~~~$N_{\rm e}$ = 700\,cm$^{-3}$ 	\\
\NII\ ($\lambda$6548+$\lambda$6584)/$\lambda$5755	       	&	~~~195:	 				&    ~~~$T_{\rm e}$ = 7600\,K:			\\
\OII\  $\lambda$3727/($\lambda$7320+$\lambda$7330)		&	~~~~~77:   				&     ~~~$T_{\rm e}$ =	6700\,K:			\\
\SIII\  ($\lambda$9069+$\lambda$9532)/$\lambda$6312		&	~~~198	   				&     ~~~$T_{\rm e}$ =	6700\,K:			\\
\OIII\  ($\lambda$4959+$\lambda$5007)/$\lambda$4363~~~  	&  $>$210:					&	~~~$T_{\rm e}$  $<$ 9950\,K		\\
\HeI\ $\lambda$7281/$\lambda$5876     					&     ~~~~~~~0.17      			&     ~~~$T_{\rm e}$   =  7300\,K		\\
\HeI\  $\lambda$7281/$\lambda$6678     					&    ~~~~~~~0.053			&     ~~~$T_{\rm e}$ =  7700\,K	    		\\
\hline
Gas temperature (adopted)								&     ~~~~~~~...				&     ~~~$T_{\rm e}$ = 7500\,K			\\
\hline		
\end{tabular}
\end{center}
}
\end{table}

\subsubsection{Expansion Velocity}\label{sec:exp_vel}

There do not appear to be any expansion velocity measurements for Abell~48 in the literature, so we used the WiFeS red-arm medium-resolution spectrum, taken in April 2010, to estimate the expansion velocity of the nebula from the gaussian profiles of the brightest nebular emission lines.  The observed full-width at half-maximum (FWHM) for each line was determined using the \textit{splot} function in {\sc IRAF}, and the expansion velocity, corrected for instrumental resolution and thermal broadening, was calculated with the following expression (Gieseking, Hippelein \& Weinberger 1986):
\begin{equation}\label{eq:fwhm}
v_{\rm exp} = 0.5\,({\rm FWHM}_{\rm obs}^{2} -  {\rm FWHM}_{\rm instr}^{2} - 8({\rm ln}\,2)k\,T_{\rm e}/m)^{0.5}
\end{equation}


where $k$ is Boltzmann's Constant, T$_{\rm e}$ is the electron temperature taken from section~\ref{sec:nebular_spectroscopy}, $m$ is the atomic mass of the measured ion, and FWHM$_{\rm instr}$  is the instrumental FWHM determined from several night-sky lines.  The HWHM is often taken to be equal to the expansion velocity (e.g. Weinberger 1989), but we instead use $V_{10}$ (Medina et al. 2006) which is a better proxy for the expansion of the outer nebular rim  (e.g. Sch\"onberner et al. 2005).  We derive $V_{10}$ = 40 \kms, which agrees with the  average expansion  velocity of sample of [WC] PNe, $v_{\rm exp}$ of 36~km~s$^{-1}$ (Pe\~na, Medina \& Stasi\'{n}ska 2003).  From the same data-set we measure a heliocentric radial velocity of +36 $\pm$ 3\,\kms, obtained using the {\sc IRAF} package {\tt emsao}, which is lower than the velocity of +50 $\pm$ 4\,\kms\ determined by TK13.  Our measurement translates to a velocity relative to the local standard of rest of $v_{\rm LSR}$ = +49 $\pm$ 3 \kms.

This observed expansion velocity also falls in the range seen for the shells around late-WN (WNL) and luminous blue variable (LBV) stars (Nota et al. 1995; Chu 2003), but we point out that no point-symmetric, moderate-density ejecta shell is known to surround any massive WNE star.  The earliest spectral type seen in a high-surface brightness coherent nebula is WN7b for PMR~5 (Frew et al., in preparation).  Earlier WN stars possess increasingly one-sided, low-density shells, fragmented by Rayleigh-Taylor instabilities and having abundances diluted in part by swept-up gas from the diffuse ISM. An example is NGC~6888 (Gruendl et al. 2000; Fern\'andez-Mart'n et al. 2012) around the WN6b(h) star WR\,136.

\subsection{Nebular Abundances}\label{sec:A48_abundances}

We derived ionic and total nebular abundances from our spectra using the \textsc{equib} code (see Wesson, Stock \& Scicluna 2012), after adopting the values of $T_e$ and $N_e$ given in Table~\ref{table:plasma}.  We determined the helium abundance from  the measured intensities of the $\lambda$5876 and $\lambda$6678 \HeI\ recombination lines using the effective recombination coefficients from Hummer \& Storey (1987).  For the metallic ions, the abundances were derived from the observed line intensities of the strong collisionally-excited lines.  We then derived the total abundances using the ionization correction factors (ICFs) derived from the expressions given in Kingsburgh \& Barlow (1994). Note that the oxygen abundance has a larger uncertainty, as the nebular $\lambda$3727 \OII\ doublet has an uncertain flux due to the high extinction. Furthermore, the neon abundance is only an indicative estimate, owing to the poor S/N ratio of the $\lambda$3869 \NeIII\ line.   The abundances and adopted ICFs are presented in Table~\ref{A48:abundances}. 
We note that TK13 derived a significantly higher electron temperature, and therefore a lower oxygen abundance than we do.  Referring to the spectrum reproduced in their Figure\,8, we see no sign of the \OIII\ $\lambda$4363 line, nor the neighbouring \HeI\ $\lambda$4471 line, which is predicted to be stronger by our photoionization model.  Therefore we prefer our own upper limit for the \OIII\ electron temperature, and place more reliance in our nebular abundances.

\begin{table}
\begin{center}
\caption{Ionic abundances, adopted ionization correction factors and total abundances for Abell~48.}
\begin{tabular}{llc}
\hline
{ Line, $\lambda$\,({\AA})}& {Ion} & { Abundance }\\
\hline   
5876, 6678 & He${}^{+}$/H${}^{+}$ &  0.129\\
     & ICF(He) & 1.00 \\
     & He/H & 0.129  \\
\hline   
6548, 6584 & N${}^{+}$/H${}^{+}$  & 2.21($-5$) \\ 
     & ICF(N) & 2.75 \\
     & N/H &  6.07($-5$)  \\
\hline
3727, 7325 & O${}^{+2}$/H${}^{+}$  & 1.76($-4$) \\ 
4957, 5007 & O${}^{+2}$/H${}^{+}$ & 3.13($-4$) \\ 
     & ICF(O) &  1.00\\
     & O/H & 4.89($-4$)   \\
\hline
3869 & Ne${}^{+2}$/H${}^{+}$ & 1.10($-4$) \\ 
     & ICF(Ne) &  1.57\\
     & Ne/H & 1.73($-4$)   \\
\hline  
6717, 6731 &  S${}^{+}$/H${}^{+}$  & 5.51($-7$) \\ 
6312, 9069~~ & S${}^{+2}$/H${}^{+}$ & 1.15($-5$) \\ 
     & ICF(S) & 1.10 \\
     & S/H & 1.33($-5$) \\
\hline  
7136 & Ar${}^{+2}$/H${}^{+}$  & 2.10($-6$) \\ 
     & ICF(Ar) & 1.57 \\
     & Ar/H &  3.30($-6$)  \\
\hline \label{A48:abundances}
\end{tabular}
\end{center}
\end{table}

Our analysis shows that Abell~48 has no significant enrichment of nitrogen (N/O $\approx$ 0.12)  and does not belong to Peimbert's Type I class (Peimbert 1978; Peimbert \& Torres-Peimbert 1983), for which we adopt Kingsburgh \& Barlow's (1994) definition of N/O $>$ 0.8.   In fact, the nebular abundances are approximately solar, a point to which we return below.  Here we differ from the conclusions of TK13, who find slightly subsolar abundances for Abell ~48.  We attribute this discrepancy to the higher electron temperature they adopt.  We do not detect the \OIII\ $\lambda$4363 line which sets an upper limit on the electron temperature of $\sim$10 kK, and the mean of several independent diagnostics suggests a lower temperature of 7500\,K (see Table~\ref{table:plasma}), and hence a higher derived oxygen abundance.  We will revisit the nebular abundances in \S~\ref{sec:neb_abundances}, \S~\ref{sec:WNclass} and \S~\ref{sec:siblings}.


\section{The Central Star}\label{sec:cspn}

\subsection{Photometry}\label{sec:photometry}

The CSPN is obvious on SuperCOSMOS $B_J$, $R_F$, $I_{N}$, and \ha\ images (Hambly et al. 2001; Parker et al. 2005), as well as 2MASS, UKIDSS and GLIMPSE images (Skrutskie et al. 2006; Lawrence et al. 2007; Benjamin et al. 2003).  We summarise the available literature photometry in Table~\ref{table:CS_phot}, along with new $UBVI$ magnitudes measured by us.   In order to correct the observed magnitudes and colours for reddening, a colour excess of E$(B-V)$ = 1.90 mag is adopted from \S\,\ref{sec:reddening}.  The visual absorption is then $A_v$ = 5.9 mag using the reddening law of Howarth (1983).

\begin{table}
{\footnotesize	
\begin{center}
\caption{Summary of photometric measurements for the [WN4-5] central star of Abell~48.  The third column gives the dereddened magnitudes.}
\label{table:CS_phot}
\begin{tabular}{clcl}
\hline
~Waveband~ & ~~~~~~m &m$_0$ & Source\\
\hline								
$U$                  &    19.8  $\pm$ 0.2    &   ~~~10.6~~~   &        This work \\
$B$                  &    19.48  $\pm$ 0.02    &   11.7   &       This work   \\
$V$                  &  17.80 $\pm$ 0.01    & 11.9      &   This work              \\
$V$                  &  17.72 $\pm$ 0.25    & 11.8      &   YB6             \\
$I$                  &  15.50 $\pm$ 0.05       &  12.0  & DENIS          \\										
$I_{c}$           &  15.14 $\pm$ 0.05       &  11.7  &This work         \\										
\hline
$J$                 & 13.54 $\pm$ 0.09       & 11.8  & DENIS     \\		
$J$                 & 13.508 $\pm$ 0.027   & 11.7  & 2MASS     \\		
$J$                 & 13.440 $\pm$ 0.002   & 11.7  & UKIDSS      \\		
\hline
$H$                & 12.834 $\pm$ 0.028   &   11.7 & 2MASS      \\					
$H$                & 12.823 $\pm$ 0.001   &   11.7 & UKIDSS      \\					
\hline
$K_{s}$         & 12.302 $\pm$ 0.11     &  11.6 & DENIS   \\
$K_{s}$         & 12.325 $\pm$ 0.027   &  11.6 & 2MASS     \\  
$K_{s}$         & 12.281 $\pm$ 0.002   &  11.6 & UKIDSS      \\
\hline
{[}3.6]            &  11.69 $\pm$ 0.06	    & 11.3  & GLIMPSE     \\ 
{[}4.5]            &  11.25 $\pm$ 0.10      &  11.0 & GLIMPSE    \\
{[}5.8]            &  11.06 $\pm$ 0.09      &  10.8 & GLIMPSE      \\
{[}8.0]            &  11.04 $\pm$ 0.16      &  10.8 &  GLIMPSE       \\
\hline
\end{tabular}
\end{center}
}
{\footnotesize References for photometry: 2MASS (Skrutskie et al. 2006);  DENIS (Epchtein et al. 1994); GLIMPSE (Benjamin et al. 2003); UKIDSS (Lawrence et al. 2007); YB6 (Zacharias et al. 1994). }
\end{table}

No time series photometry of the CSPN is available, so information on any short-period variability is lacking.  However $J$ and $K_s$ magnitudes are available from three different epoch surveys, DENIS, 2MASS and UKIDSS (Epchtein et al. 1994; Skrutskie et al. 2006; Lawrence et al. 2007) and all agree within the uncertainties, so we can tentatively say the star is not a large amplitude variable, unlike the ionizing star of the LMC planetary N\,66 (McKibben Nail \& Shapley 1955), which is likely to host an interacting binary system (Hamann et al. 2003).

\subsection{Spectroscopy}

Spectroscopic classifications of the central star of Abell 48 were undertaken by Wachter et al. (2010) and DePew et al. (2011).  Wachter et al. (2010)  classified the central star as a WN6 according to the classification scheme developed by Conti, Massey \& Vreux (1990).  The previous shallow MSSSO~2.3m spectrum presented in DePew et al. (2011) showed a broad feature at $\lambda$7116\AA\ in the red, but a large section of the spectrum between 5050\AA\ and 6300\AA\ was not observed, so it was unclear whether this was a [WN] or [WN/C] star.  
In our deeper spectra, we again find a strong feature centred around $\lambda$7116\AA, attributed to the complex of \NIV\ lines between $\lambda$7103\AA\ and $\lambda$7129\AA.  The \HeII\ $\lambda$6560 line appears as a broad (FWHM$\sim$25\AA) feature from which the nebular H$\alpha$ line protrudes (the nebular lines are over-subtracted in the Palomar data).  The \HeII\ $\lambda$4686 and $\lambda$5412 lines are prominent and a broad \CIV\ $\lambda\lambda$5801,12 doublet is also present, as is typical of the WN class.

\begin{table}
\begin{center}
\caption{Principal emission lines found in the CSPN of Abell 48. The line equivalent widths and uncertainties are in \AA. }
\label{line_table1}
\begin{tabular}{lcccc  }
\hline
Species & $\lambda$& $-W_{\lambda}^a$  & $-W_{\lambda}^b$ & FWHM    \\
                &  (\AA)        & (\AA)                           &(\AA)                   &  (km s$^{-1}$)          \\ [0.5ex]
\hline
\NIV\ 	& 3480 & ...  & 45 $\pm$ 10  & 1300:    \\
\NIV\ 	& 3748 & ...  &   3.0 $\pm$ 1.0  & ...           \\
\NIV\ 	& 4058 & ...  & 23 $\pm$ 3  & 830       \\
\HeII\,+\,\NIII\ & 4100 &  ... & 29 $\pm$ 4  &   1060      \\ 
\HeII\,+\,\NIII\ & 4200 & ...  & 15 $\pm$ 2  &   800       \\
\HeII 	& 4339 & ...  &  11.5 $\pm$ 2.0 &   800      \\
\NIII 	& 4379 &  ... &  6.0 $\pm$ 1.0 &   ...      \\
\NIII 	&   4518 & ...  & 12 $\pm$ 2.5   & ...  \\
\HeII 	&  4542& n.m. &  16 $\pm$ 4  & 1200 \\
\NV 		&  4604 & n.m. & 22.5 $\pm$ 3.0  & 560   \\
\NV 		&   4620 &  n.m.   & 19.5 $\pm$ 3.0  & 600    \\
\NIII 	&   4634-40 & ...  & 12 $\pm$ 3 & ...   \\
\HeII 	& 4686 & ...  &  160 $\pm$ 20 & 1170    \\
\HeII 	& 4860 &  ... & 21.5 $\pm$ 4.5  & 1120   \\
\NIII 	&   4905 & ...  & 2.0 $\pm$ 1.0  & ...   \\
\NV 		&   4944 & n.m.  & 10.0 $\pm$ 3.0  & ...   \\ 
\NIV 	& 5202 & 5.4  $\pm$ 2.0  &  5.0 $\pm$ 1.4 & ...           \\
\HeII 	&  5411 &  23 $\pm$ 4 &  34 $\pm$ 6 & 900    \\ 
\CIV 	& 5806  &  ...  &  15.2 $\pm$ 3.0  & 1300: \\
\HeI   	& 5876  & 3.8 $\pm$ 2.0   & n.m.  & ...    \\
\HeII     	& 6118 &  4.0 $\pm$ 1.4 & 2.2 $\pm$ 1.0      \\ 
\HeII  	& 6311 & 8.0 $\pm$ 2.0  & 5.4$\pm$ 1.6   & ...   \\ %
\NIV   	& 6382 & 3.8 $\pm$ 2.0  &   6.8 $\pm$ 2.0  & 800    \\
\HeII  	& 6406 & 7.1 $\pm$ 1.0  &  7.9 $\pm$ 1.0  & ...   \\  
\HeII  	& 6527 & 4.3 $\pm$ 1.0  &  6.6 $\pm$ 1.0  & ...    \\ 
\HeII   	& 6560 & 130 $\pm$ 30  & n.m.  & 1020   \\  
\HeI\,+\,\HeII&  6683 & ...  &  ...  & blend  \\  
\HeII   	& 6891  & 11.4 $\pm$ 1.0   & 10.4 $\pm$ 1.0 &  750       \\  
\NIV  	& 7116 &  160 $\pm$ 20 & 130 $\pm$ 10 & 1300  \\
\HeII 	& 7178 &  16.7$\pm$ 3.2 & ...  & ...    \\ 
\HeII\,+\,\NIV  & 7590& n.m. & 9.8$\pm$ 2.0  & 450:  \\ 
\HeII 	& 8237  & 40.3 $\pm$ 5.3  &  31.1 $\pm$ 3.0 & 850   \\  
\HeII		& 9225  & 7.9$\pm$ 1.8  & ...  & 700   \\  %
\hline
\end{tabular}
\begin{flushleft}
\footnotesize{$^{a}$ Second WiFeS spectrum;  $^{b}$ Hale spectrum; n.m. = detected but not measured.
}
\end{flushleft}
\end{center}
\end{table}

We measured the stellar emission lines from all of our spectra using the {\tt splot} function in {\sc IRAF}.   Table~\ref{line_table1} presents the identifications, the line equivalent widths (in \AA), and the line widths (FWHM in \kms) of the principal lines that were clearly detected.  The typical uncertainties are 10--15\% (no less than $\pm$ 1\AA\ for the weak lines) and $\pm$\,50\,\kms\ respectively, based on repeat measurements.  
The numerical criteria we used to classify the star are summarised in Table~\ref{table:classify}.  Besides the high S/N Palomar spectrum, we also made use of our low-resolution WiFeS stellar spectrum, which extends to $\sim$9500\AA, but very few classification schemes use the far-red / NIR wavelength region.  As part of another project, we downloaded all the red spectra of the Galactic WN stars presented by Hamann, Koesterke \& Wessolowski (1995), and measured the equivalent width of the $\lambda$8237 \HeII\ line  for each star, if available.  We tabulated the \HeII/\HeI\ ratios and spectral type from Smith, Shara \& Moffat (1996) for each and found, as expected, a tight relation between the two quantities. This  suggests a spectral class of [WN4] for the CSPN of Abell 48.   In summary, the \NIV $\lambda$4057 and \NV $\lambda\lambda$4604,20 lines being much stronger than the \NIII $\lambda$4634,40 blend, with relatively weak \CIV, we revisited the classification and now find an earlier type than [WN6] (c.f. Wachter et al. 2010), after primarily following the criteria of Smith et al. (1996).  By giving most weight to the ratio of the equivalent widths of the \HeII $\lambda$5411 and \HeI $\lambda$5876 lines, our  adopted classification is [WN4--5], while TK13 estimated a spectral class of [WN5].

\begin{figure*}\label{fig:CSPN_spectrum}
\begin{center}
\includegraphics[width = 17cm]{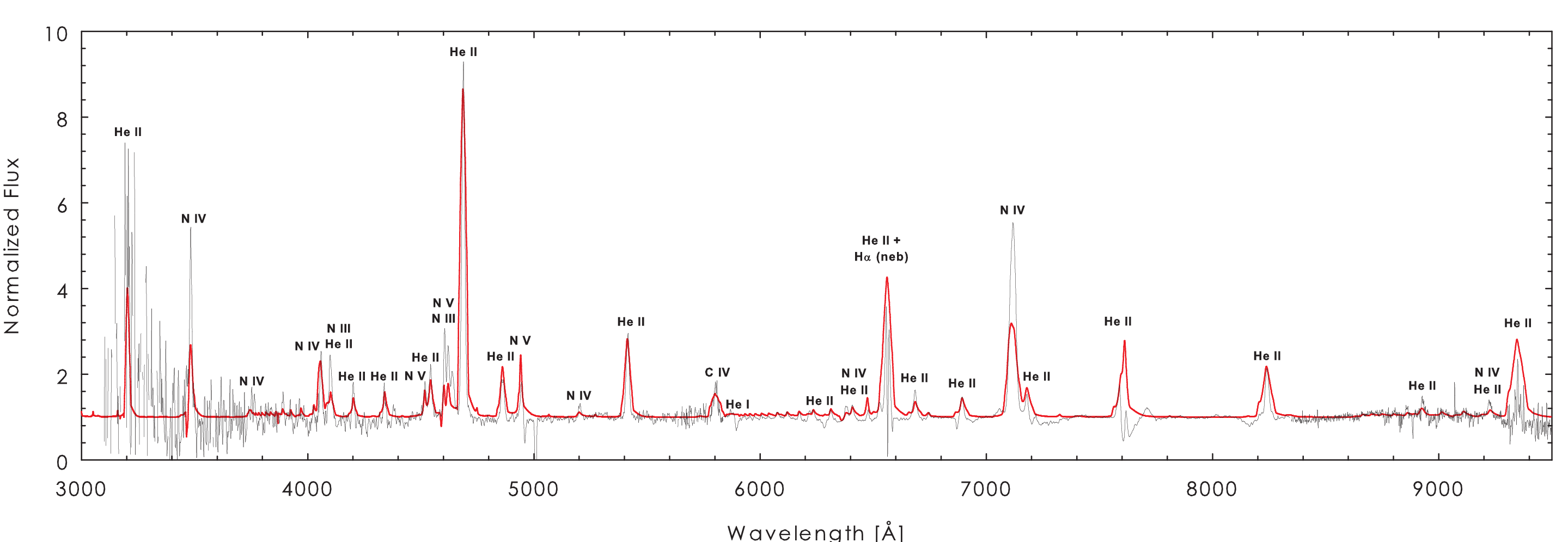}
\caption{Normalized spectrum (black line) of the CSPN, spliced from our WiFeS and Palomar data. Note the presence of the \NV\ $\lambda\lambda$4604,4620 lines and the very prominent \HeII\ $\lambda$4686 and \NIV\ $\lambda$7116 features in the spectrum.  Overplotted is an updated model 10-13 (red line) from the Potsdam Wolf-Rayet (PoWR) grid, (provided by H. Todt), which fits the relative strengths of the He lines well.  The relatively poor fit to the nitrogen lines suggests the default nitrogen abundance is too low.  A colour version of this figure is available in the online journal. }
\end{center}
\end{figure*}

\begin{table}
{\footnotesize	
\begin{center}
\caption{Classification of the central star, after Smith et al. (1996). The second column is the ratio of the measured equivalent widths. }
\label{table:classify}
\begin{tabular}{lcc}
\hline
Criterion    								& Ratio 		 & Classification    \\
\hline
$W_{\lambda5411}/W_{\lambda5876}$    	&  9.5		 &     WN4      	 \\
$W_{\lambda8237}/W_{\lambda5876}$    	&  7.8		 &     WN4      	 \\
$P_{5411}/P_{5876}$    					&  2.6		 &     WN5      	 \\
$P_{4604}/P_{4640}$    					&  1.6		 &     WN5    	 	 \\
$P_{4057}/P_{4604-40}$   				&  0.8		 &     WN4    	 	 \\
$P_{5808}/P_{5876}$        				&  1.6		 &     WN5     	 	 \\
$P_{5808}/P_{5411}$              			&  0.6		 &     WN4--5     	 \\
\hline
Adopted   							& ...			&     [WN4--5]        \\
\hline
\end{tabular}
\end{center}
}
\end{table}

We further note that there is no obvious indication of an oscillating Pickering decrement (e.g. Smith et al. 1996), showing that the star is hydrogen deficient.  To quantitatively estimate the hydrogen fraction, we use the `Pickering index' defined by  Smith et al. (1996; cf. Oliveira, Steiner \& Cieslinski 2003):
\begin{equation}
\frac{N({\rm H}^{+})}{N({\rm He}^{++})} = \frac{W(\lambda4859+\lambda4861)}{[W(\lambda4541) \times W(\lambda5411)]^{0.5}} - 1
\end{equation}

Formally, we obtain the unphysical value of $-0.1$, which we reset to zero, on the basis of the observed uncertainties of the equivalent widths.  In practice, we adopt a working upper limit for the hydrogen abundance of $\sim$10\% based on the observed S/N ratio of the lines.   Overall the spectrum is similar to a Population I WN4o star (Vreux, Dennefeld \& Andrillat 1983; Smith et al. 1996), but as we show in Section~\ref{sec:interpretation}, the  properties of the surrounding nebula unambiguously show that we are dealing with low-mass CSPN.

\subsection{Model Atmosphere Analysis}\label{sec:model}

To begin the analysis, we downloaded the synthetic spectra appropriate for WNE model atmospheres from the Potsdam Wolf-Rayet (PoWR) database\footnote{\url{http://www.astro.physik.uni-potsdam.de/~wrh/PoWR/powrgrid1.html}} (Hamann \& Gr\"afener 2004) to compare with our data.   These spectra are determined from the PoWR model atmospheres, described in Gr\"afener, Koesterke \& Hamann (2002) and Hamann \& Gr\"afener (2003), which account for spherical expansion, non-LTE effects, and metal line blanketing.  The PoWR model grid has two dimensions, the stellar temperature $T_\star$ and the ``transformed radius,'' $R_{\rm t}$  (Hamann \& Koesterke 1998), which is a wind density parameter, being a function of the mass-loss rate, $\dot{M}$, and the stellar radius, $R_{\star}$.  It is given by:
\begin{equation}\label{eq:transform}
R_{\rm t} = R_{\star} \left[   \frac{v_\infty}{2500~{\rm km\,s^{-1}}} \Big{/}   \frac{\dot{M}\,\sqrt{D}}{10^{-4}~M_{\odot}\,{\rm yr^{-1}}}  \right]^{2/3}
\end{equation}
where $v_\infty$ is the terminal velocity of the wind and $D$ is a wind clumping factor (or inverse of the filling factor).  It has been previously noted that many parameters of massive WRs and [WR]s are similar (e.g. Crowther, Morris \& Smith 2006b), a consequence of the scaling law for WR atmospheres first noted by Schmutz, Hamann \& Wessolowski (1989).   In other words, two atmospheres with the same temperature ($T_\star$) and transformed radius ($R_{\rm t}$) show very similar spectra regardless of $L_\star$, $R_\star$, $\dot{M}$ and $v_{\infty}$.

The PoWR grid models are normalized to $L/L_\odot$ = 2.0$\times10^5$ which is appropriate for a massive WN star; consequently, the stellar flux of our CSPN is  proportional to $L_{\star}$, and $R_{\star}$ and $\dot{M}$ are proportional to $L_{\star}^{1/2}$ and $L_{\star}^{3/4}$ respectively.  
We follow Hamann \& Koesterke (1998) in assuming a wind clumping factor of $D$ = 4 ($f$ = 0.25), which is appropriate for WN stars.
The stellar radius, $R_{\star}$, is the inner boundary of the model atmosphere and corresponds by definition to a Rosseland optical depth of 20. The stellar effective temperature, $T_{\star}$, is the effective temperature at $R_{\star}$, and can be easily calculated from the Stefan-Boltzmann law:
\begin{equation}\label{eq:stefan}
L = 4 \pi R_\star^2 \sigma T_\star^4
\end{equation}
For Abell~48, various line ratios using the \NIII, \NIV\ and \NV, and \HeI\ and \HeII\ lines, initially constrained the temperature to between 63 and 79 kK and log $(R_{t}/R_{\odot}$) between 0.6 and 1.0.  The absence of \HeII $\lambda4686$ emission in the surrounding nebula is due to the optically thick WR wind.  Models with $T_{\rm eff} <$ 60kK show a \HeI\ $\lambda$5876 line much stronger than observed, while models with $T_{\rm eff} >$ 90\,kK are ruled out as these produce significant flux shortward of the HeII ionization edge at 228\,\AA, so \HeII\ emission should be clearly present in the surrounding nebula.  The estimated temperature is in broad agreement with the temperature range of  60--90\,kK  found for massive WN4 stars by  Hamann \& Gr\"afener (2003).

The best-fit model  has $T_{\rm eff}$ = 71 kK and $R_{\rm t}$ = 6.3 $R_{\odot}$ (model 10-13) and reproduces fairly well the relative equivalent widths of the helium lines.  A new  synthetic spectrum calculated from a revised model 10-13, which utilised  updated atomic data was kindly provided by H. Todt (2013, pers. comm.)  The model parameters are $T_{\rm eff}$ = 71kK and log$R_t$ = 0.8, with the default abundances set to He:C:N:O = 0.98: 1E-4: 0.015: 0.0. The observed and synthetic spectra are plotted in Figure~\ref{fig:CSPN_spectrum}).  However, the \NIII, \NIV\ and \NV\ line intensities are all weaker in the model, in particular the \NV\ 4944 and \NIV\ 7116 blends.   This indicates that the default nitrogen abundance is too low.  We have not convincingly detected any line due to oxygen, so we make no modification to the default model abundance.  Our parameters are in excellent agreement with that derived by TK13 (see their Table~3), the main difference is that TK13 estimate a higher nitrogen abundance of 5 percent and a hydrogen abundance of 10 per cent, a result that does not disagree with our upper limit for this element.  So it is now apparent that the photospheric nitrogen and hydrogen abundances differ considerably between Abell~48 and IC~4663, mimicking the range of H abundances seen in the O(He) stars (Reindl et al. 2013, and references therein).


The intrinsic luminosity of the star, and hence its distance, is not constrained from the model atmosphere analysis. Instead we adopt the distance of 1.6\,kpc from \S~\ref{sec:distance} in the discussion that follows, cf. TK13 who assumed a canonical post-AGB luminosity.  We found the bolometric correction of the model atmosphere to be $-5.4$ mag, which leads to an estimated luminosity of the star of $\sim$5.5 $\times 10^{3} L_{\odot}$.
From our adopted value of the temperature $T_\star$, the stellar radius $R_\star$ was calculated using equation~\ref{eq:stefan} to be 0.49\,$R_{\odot}$.  From the line widths presented in Table~\ref{line_table1}, we estimate a terminal wind velocity, $v_{\infty}$ of 1200 \kms.  Finally, we estimate the mass-loss rate, $\dot{M}$, which is proportional to $L^{3/4}$ for a constant $R_{\rm t}$ (equation~\ref{eq:transform}).  We determine log\,$\dot{M}$ = $-6.3\pm 0.2$ $M_{\odot}$ yr$^{-1}$.

\subsection{Stellar Spectral Energy Distribution}\label{sec:stellar_SED}

\begin{figure}
\begin{center}
\includegraphics[width=8.6cm]{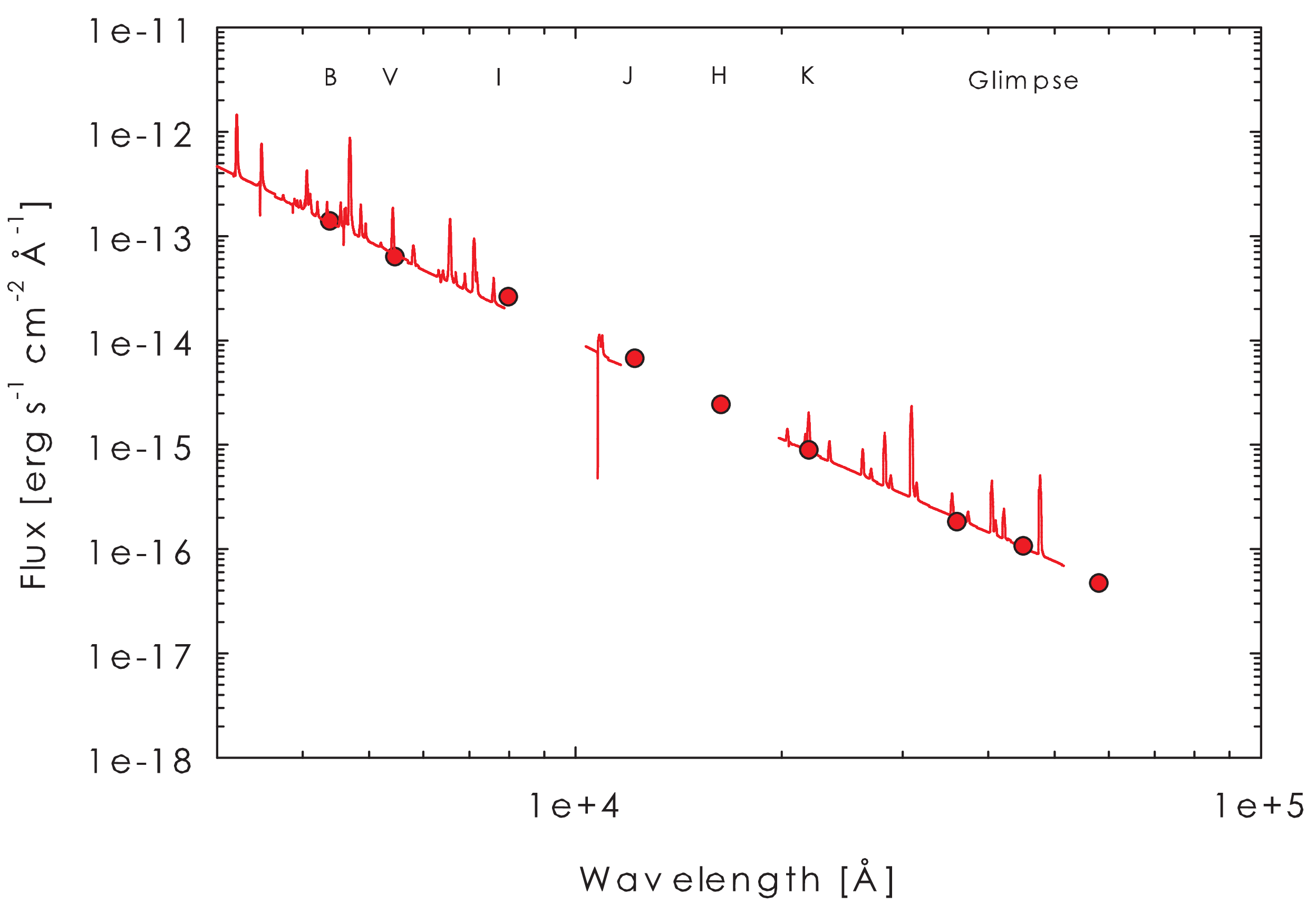}
\caption{The SED of the central star of Abell 48.  The red line shows the synthetic spectrum (PoWR model 10--13), with the photometric measurements dereddened with $E(B-V)$ = 1.90 plotted as red circles.  A colour version of this figure is available in the online journal. }
\label{SED}
\end{center}
\end{figure}

The stellar spectral energy distribution (SED) for the CSPN of Abell~48 is shown in Figure \ref{SED}, along with the adopted PoWR synthetic spectrum.  The model has been normalized to the ensemble of the dereddened values from Table~\ref{table:CS_phot},  excluding the Johnson $U$ and IRAC\,8.0\,$\mu$m magnitudes, which are both uncertain.  We applied the optical and IR extinction laws of Howarth (1983) and Indebetouw et al. (2005) respectively, using $E(B-V)$ = 1.90.  

There is no evidence from our spectra for any signature of a companion star, nor from the SED for any NIR excess due to a cooler companion (Douchin et al. 2012; De Marco et al. 2013), though in this case any limit is not strong.  The [WN] star is relatively luminous and the NIR magnitudes include a non-negligible contribution from wind free-free emission.  As a result, any companion later than $\sim$F0\,V would not be detected.  The tentative absence of variability (\S\ref{sec:photometry}) suggests the star does not have an irradiated companion (De Marco, Hillwig \& Smith 2008; Miszalski et al. 2009; Hajduk, Zilstra \& Gesicki 2010), so time-series spectroscopic analysis is probably the best way to determine if Abell~48 is a binary (Jorissen \& Frankowski 2008).  However, at orbital periods greater than a year or two, the expected velocity shifts of the emission lines would be likely too small to be detected against the stochastic variations of the stellar wind.

\section{The nature of Abell~48}\label{sec:interpretation}

As a precursor to any discussion on the evolutionary implications of the [WN] class, it is important to try to properly assess all the observational evidence for Abell~48 so that its status as a true PN or a circumstellar nebula around a massive star can be made clear.  Circumstellar nebulae can be morphologically similar to PNe, and are found around several types of massive stars (Chu 2003; FP10; Mizuno et al. 2010; Wachter et al. 2010; Gvaramadze, Kniazev \& Fabrika 2010; Boji\v{c}i\'c et al. 2011a).

\subsection{Diagnostic plots}\label{sec:diagnostics}

\begin{figure*}
\begin{center}
\includegraphics[width=8.35cm]{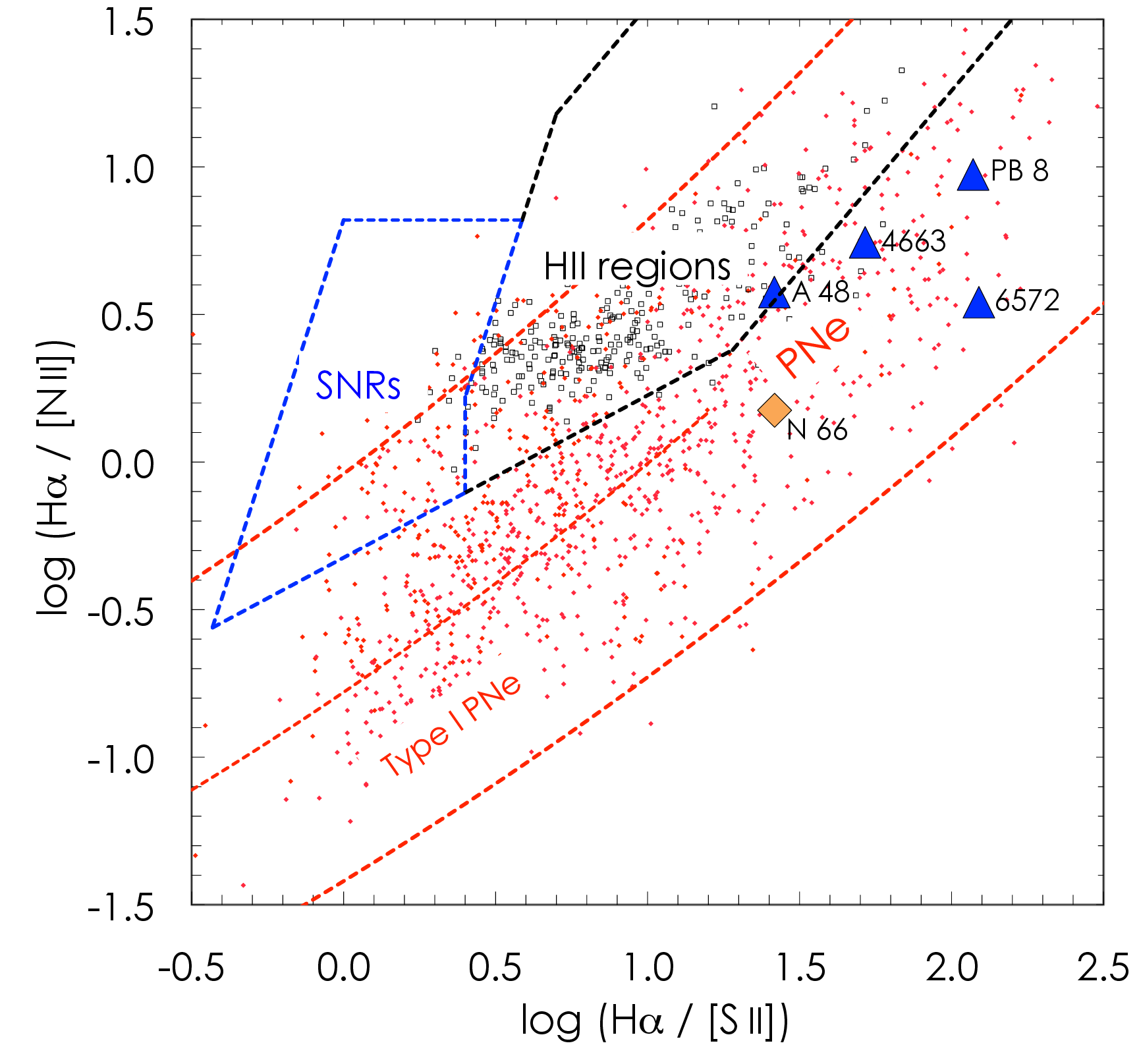}   
\includegraphics[width=8.35cm]{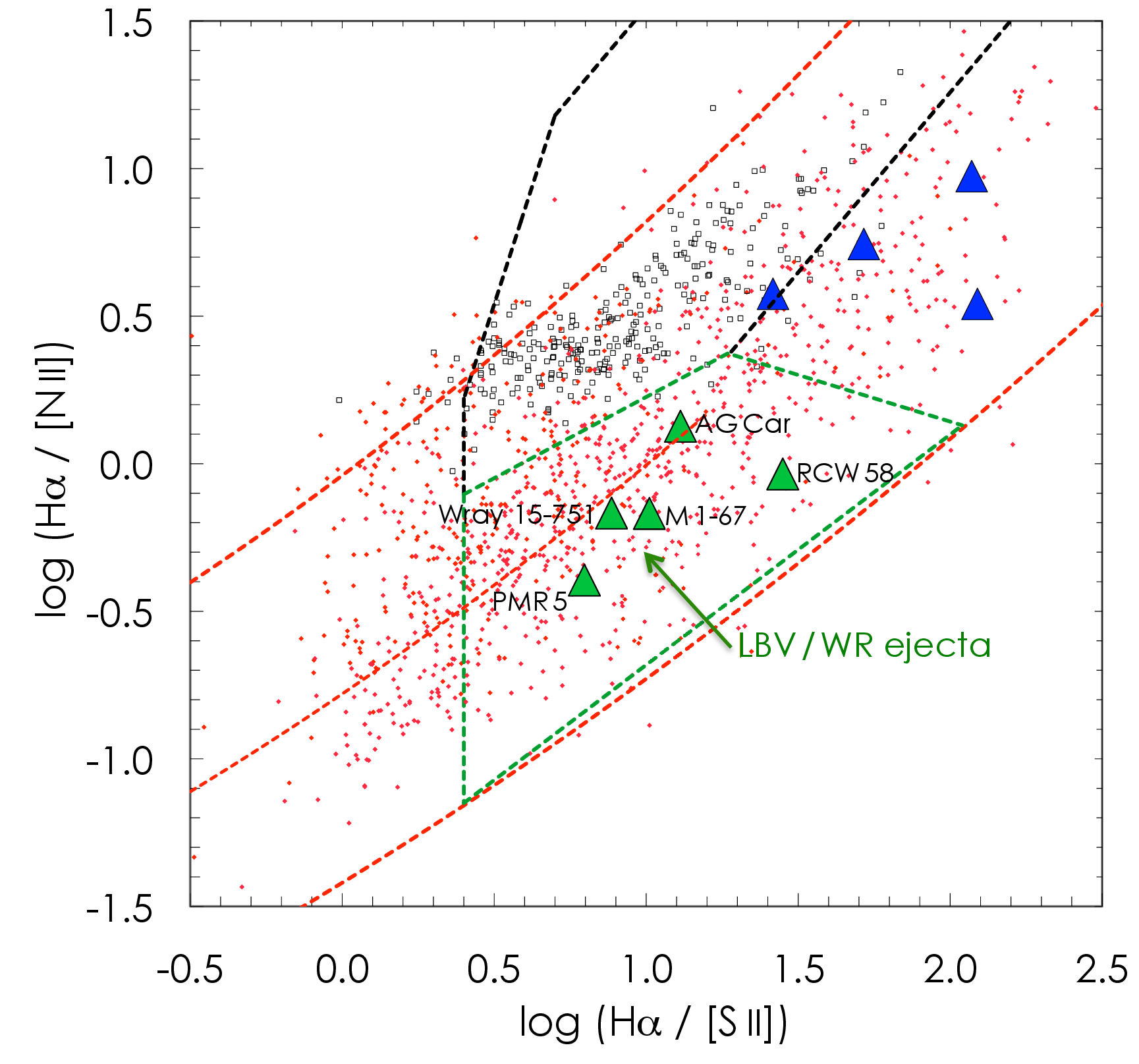}
\caption{SMB diagnostic plots updated from FP10.  Individual PNe are shown as small red dots and Galactic \HII\ regions as small black squares.  In the left panel the regions populated by HII regions, SNRs and PNe are marked; note the considerable overlap between them. The blue triangles show Abell 48, IC 4663, PB 8,  and NGC 6572, which has a candidate [WN/C] CSPN (Todt et al. 2012).  These all plot squarely in the PN domain, while the orange diamond is the LMC planetary N\,66.  In the right panel the domain of LBV/WR ejecta is plotted, and the green triangles plot a selection of ejecta nebulae.  A colour version of this figure is available in the online journal. }
\label{fig:diag_plot}
\end{center}
\end{figure*}

In Figure\,\ref{fig:diag_plot} we present the `SMB' or log\,$F$(\ha)/$F$\,\NII\ versus log\,$F$(\ha)/$F$\,\SII\ diagnostic diagram of Sabbadin, Minello \& Bianchini (1977), as extensively updated by Frew \& Parker (2010, hereafter FP10), and further refined here (see also Sabin et al. 2013).  
Individual PNe are shown as small red dots and Galactic \HII\ regions as small black squares.  In the left panel the fields showing \HII\ regions, SNRs and PNe are marked; note the considerable overlap between them. The blue triangles show the PNe with confirmed [WN] or [WN/C] stars,  Abell 48, IC 4663, PB 8,  along with NGC 6572, which has a candidate [WN/C] CSPN (Todt et al. 2012).  These all plot squarely in the PN domain, while the orange diamond is N\,66, the only PN in the LMC with a [WN] ionizing star.  It shows substantial nitrogen enrichment compared to the mean LMC abundance.  

The right panel shows the domain of LBV/WR ejecta, and the green triangles represent several ejecta nebulae based on literature data.  These are all nitrogen enriched from CNO cycling, and mostly overplot the Type~I PN region defined by FP10.  Clearly, Abell 48 falls in the PN domain away from both Type I PNe and and the CNO processed 
ejecta around LBV and WNL stars.  This confirms our abundance analysis (\S\ref{sec:A48_abundances} and \S\ref{sec:neb_abundances}) showing the  nebula is not strongly nitrogen enriched, and  unlikely to be produced by a massive star.  
While this part of the SMB diagram has extensive overlap between PNe and \HII\ regions, Abell 48 is plainly not a \HII\ region, as its morphology and MIR colours reveal.  In particular the flux ratios, $F_{12}/F_{8}$ = 4.9, $F_{24}/F_{8}$ = 5.5, and $F_{70}/F_{12}$ = 9.2 (see Table\,\ref{table:continuum}), strongly indicate a PN rather than a \HII\ region (Anderson et al. 2012).

\subsection{Nebular Abundances}\label{sec:neb_abundances}

In Table~\ref{tab:nebabund}, we summarise the elemental abundances for the four known PNe with confirmed [WN] or [WN]/[WC] central stars, compared with a selection of ring nebulae around massive stars. 
As noted above, Abell~48 appears to have no significant nitrogen enrichment, and modest He enrichment, so it does not belong to Peimbert's Type I class (Peimbert 1978; Kingsburgh \& Barlow's 1994).   We use the nebular argon abundance as a metallicity indicator, as its abundance is predicted to be unaltered by nuclear processes in PN progenitor stars.  We conclude  that the progenitor of Abell~48 was slightly metal rich with [Ar/H] = +0.12 dex, but owing to the uncertainty of the model-dependent ICF for argon (only one ionization stage is observed), Abell\,48 is consistent with having a solar argon abundance.  The metal abundances of  PB~8 (Todt et al. 2010a) and IC 4663 (MC12) are similar, being approximately solar, or perhaps slightly more metal rich.  We also note that the evolved PNe K~1-27 and LoTr~4, around O(He) stars, also show only modest helium enrichment, but the detailed CNO abundance pattern is difficult to determine for these very high excitation, optically thin PNe (Rauch, Koppen \& Werner 1994, 1996).

Also recall that strongly enhanced nitrogen and helium abundances are expected, and seen, in the ejecta around massive LBV  and WNL stars that are undiluted by swept-up ISM (Dufour 1989; Kwitter 1981, 1984;  Esteban et al. 1992; Smith et al. 1994; Lamers et al. 2001; Stock, Barlow \& Wesson 2011).  LBV and WR ejecta also show evidence of oxygen depletion (e.g. Dufour 1989), as seen in Table~\ref{tab:nebabund}, but the oxygen abundance of Abell\,48 is very close to solar.   
The  N/O mass ratio is high in LBV and WNL ejecta, ranging up to $\sim$200 for the ejecta around the peculiar WR-like star NaSt~1 (Crowther \& Smith 1999), while the value for Abell~48 is only 0.12, rather typical of non-Type I PNe (Kingsburgh \& Barlow 1994).

\begin{table*}
\caption{Elemental abundances, normalized to log(H) = 12.0, for the three confirmed Galactic PNe with [WN] or [WN/WC] central stars, and N\,66 in the LMC, compared to a sample of ejecta-dominated nebulae around massive stars.  The reference sources are given in the footnotes to the table. For comparison, we also list the average abundances for Type\,I and non-Type\,I PNe, the abundances of the Orion nebula, and the solar abundances from Asplund et al. (2009).}
\label{tab:nebabund}
\begin{center}
\begin{tabular}{l l l l r c c c c c c c r c c l}
\hline
Name		  	&  $|$ &   SpT       &  $|$ &      He~~  	&         N   	  &     O    		&   Ne      		&     S      	  &     Ar   	 &  $|$  &  N/O	& [O/H]	    &  [Ar/H]& $|$  &  Reference   \\ 
\hline 
Abell 48     	       &  $|$  &   [WN4--5]     	&  $|$ & 	11.11  	& 	7.78   &   8.69		&    8.2:       	&  7.12         &    6.52	  & $|$  &   0.12  & $+0.00$	& $+0.12$     &  $|$  &      This work 	 \\ 
IC 4663		       &  $|$  &   [WN3]     		&  $|$ & 	11.11  	& 	8.26   &   8.70		&    8.11       	&  7.02        &    6.43	  & $|$  &   0.36   & +0.01 &  $+0.03$    	    &  $|$  &      MC12		 \\
PB 8		     	&  $|$   & [WN6/C7] 	&  $|$ &       11.09  	& 	8.21   &   8.76		&    8.13        	&  7.31        &    6.64	  & $|$  &    0.28& +0.07 &  $+0.24$    	       & $|$  &       GR09		 \\ 
LMC-N66   	       &  $|$   &  [WN4.5]  		&  $|$ &       11.04  	& 	7.97   &   8.37		&    7.93        	&  6.68        &    6.11	  & $|$  &   0.40& $-0.32$ & $-0.29$       & $|$  &       P95,T03,B04		 \\ 
\hline
NaSt~1     	  	&  $|$   &    pec 	&  $|$ & 	...~~	  	& 	8.87  &   6.56		&   7.89        		&  6.94   & 6.43	  & $|$  &  200   & $-2.13$	      &  $+0.03$   & $|$  &        CS99		 \\ 
$\eta$ Car    	       &  $|$   &      LBV    	 &  $|$ & 	11.26 	& 	9.04  &   7.29:        &   7.98     	&  (7.1)	           & (6.4)		  &  $|$   & 56:   & $-1.4:$ &  (0.0)   & $|$  &      DG97		\\ 
AG Car     	  	&  $|$   &    WN11h 	  	&  $|$ & 	...~~	  	& 	8.22   &   7.52		&    ...        		&  $>$6.7   & ...		  & $|$  &  5.0   & $-1.17$	      &  ...   & $|$  &        SS97		 \\ 
Hen 3-519	  	&  $|$   &    WN11h   	&  $|$ & 	...~~		  & 	8.21	 &   7.57		&    ...        		&  ...	           & ...		  &  $|$  & 4.4    & $-1.12$	      &   ...   &$|$  &        S97	 \\
Wray 15-751   	&  $|$   &    WN11h   	&  $|$ & 	...~~		  & 	8.6:	 &  $<$8.4	&    ...        		&  ...	           & ...		  & $|$  & $>$1.6 & $<$$-0.3$       & ...    & $|$  &        GL98	 	\\
M 1-67     	  	&  $|$   &      WN8h   	 &  $|$ & 	11.18	 & 	8.45   &   7.98		&    ...        		&  6.96	  & ...		  &  $|$  &  3.0& $-0.71$&      	...         &$|$  &        EV91	 \\ 
NGC 6888	       &  $|$    &      WN6b(h)    &  $|$ & 	11.27	  & 	8.41   &   8.14		&    ...        		&  7.18        & ...		  &  $|$  &  1.9    & $-0.55$	&     ...     &$|$  &      EV92		 \\ 
NGC 6164-5	  	&  $|$ &     O6.5f?p     &  $|$ & 	$>$11.11	  & 	8.13   &   8.25	         &    7.5:        	&  7.1          & $>$6.3   & $|$ &  0.76    & $-0.44$      &  $>$$-0.1$   &$|$  &     DP88		\\ 
\hline 
Type\,I PN         	&  $|$   &     ...          	&  $|$ & 	11.11  	&      8.72   &   8.65		&    8.09      	&  6.91        &    6.42	  & $|$  & 1.17   & $-0.04$	&   $+0.02$          & $|$  &       KB94		  \\ 
non-Type\,I PN~~~&  $|$  &     ...          		&  $|$ & 	11.05	&      8.14    &   8.69		&    8.10      	&  6.91	     &    	6.38    & $|$   & 0.28    & +0.00	 &    $-0.02$      & $|$  &      KB94		  \\ 
Solar     	           	&  $|$ &     ...        	   	&  $|$ & 	10.93  	&      7.83   &   8.69		&    7.93      	&  7.12        &    6.40	   & $|$  & 0.14  & 0.00	 &    ~~0.00        & $|$  &        AG09	           \\ 
Orion	             &  $|$ 	&     ...         	 	&  $|$ & 	10.99	&      7.78   &   8.67  		&    8.05	   	&  7.08       &    	6.49    & $|$  &  0.13 & +0.04	&  $+0.22$           &  $|$  &       D84, EP04		  \\ 
LMC	             &  $|$ 	&     ...         	 	&  $|$ & 	10.99	&      7.14   &   8.35  		&   7.61	   	&  6.70       &    	6.29    & $|$  &  0.06 & $-0.32$	&  $-0.11$           &  $|$  &       RD92		  \\ 
\hline
\end{tabular}
\end{center}
\begin{flushleft}
{\footnotesize References: ~~AG09 -- Asplund et al. (2009); B04 - Bernard-Salas et al. (2004); CS99 -- Crowther \& Smith (1999); D84 -- Dufour (1984);  DP88 -- Dufour et al. (1988); DG97 -- Dufour et al. (1997); EP04 -- Esteban et al. (2004); EV91 -- Esteban et al. (1991); EV92 -- Esteban \& Vilchez (1992);  GL98 -- Garcia-Lario et al. (1998); GR09 -- Garc\'ia-Rojas et al. (2009); KB94 -- Kingsburgh \& Barlow (1994); MC12 -- Miszalski et al. (2012b); P95 -- Pe\~na et al. (1995); RD92 -- Russell \& Dopita (1992); S97 -- Stroud (1997); SS97 -- Smith et al. (1997); T03 -- Tsamis et al. (2003).}
\end{flushleft}
\end{table*}


\subsection{Distance}\label{sec:distance}
As the distance is critical to the interpretation of Abell~48, we revisit this problem here.   Wachter et al. (2010) utilized the IRAC and 2MASS photometry to determine a large distance of 16.5 kpc on the assumption of Population\,I status, placing it on the far side of the Galactic disk beyond the bulge.  To refine this estimate, we adopt the absolute magnitude calibration for WR stars from Crowther et al. (2006a).  We average the $K_s$-band absolute magnitudes for the WN\,3-4 and WN\,5-6 subgroups from their Table\,A1 to adopt $M_{K_{s}}$ = $-3.8\pm0.7$.  Our dereddened $K_s$ magnitude from Table\,\ref{table:CS_phot} leads to a distance modulus of 15.4 mag, or a distance of 12.0$^{+4.6}_{-3.3}$\,kpc, again placing it on the far side of the Galactic bar.  The luminosity follows from the distance, reddening and bolometric correction (\S\ref{sec:model}), and is $L \approx$\,3.4 $\times 10^5 L_{\odot}$.

\begin{figure}
\begin{center}
\includegraphics[width=7.5cm]{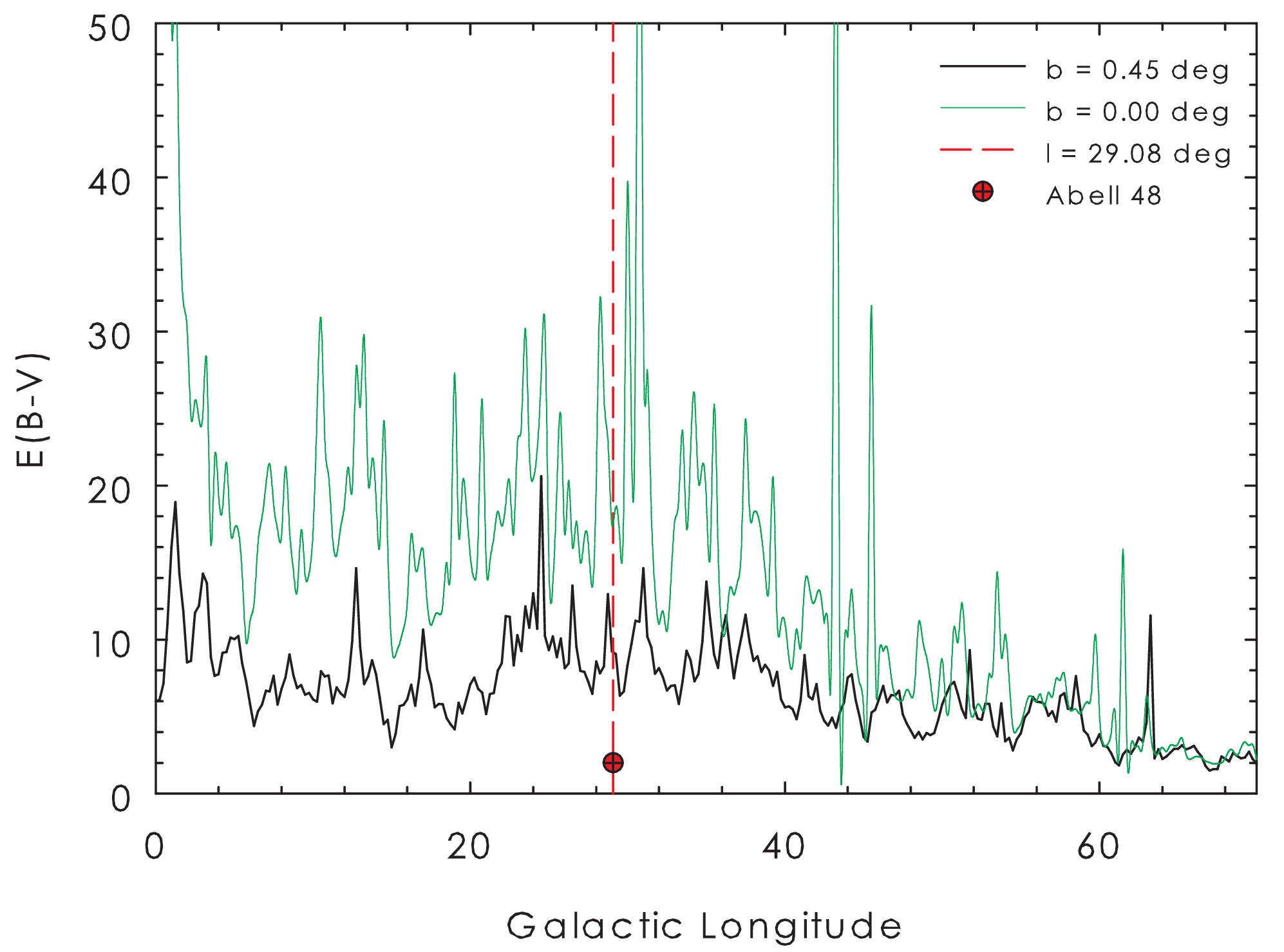}
\caption{The distribution of reddening with Galactic longitude (Schlafly \& Finkbeiner 2011), plotted for two Galactic latitudes, $b$ = 0 (the mid-plane) and $b$ = 0.45 (the latitude of Abell~48).  It can be clearly seen that the reddening to Abell 48 is a small fraction of the asymptotic reddening on that sightline, precluding any location for the nebula beyond the Galactic bar.   }
\label{fig:steps}
\end{center}
\end{figure}

The asymptotic reddening in this direction from Schlafly \& Finkbeiner (2011, who updated Schlegel et al. 1998) is $E(B-V)$ = 10.0 mag (A$_{V}\simeq$ 31 mag), compared to our adopted $E(B-V)$ of 1.90 (A$_{V}$ = 5.9).  Even though the asymptotic extinction is not very reliable in this case, there should be at least 15--20 mag of visual extinction if the star is really at or on the far side of the end of the Galactic bar.   Hence the discrepancy of such a large distance with a relatively low reddening is difficult to explain if we are dealing with a massive star, a point made by Wachter et al. (2010).
The available data strongly suggest the star is  closer and that it is  likely to be the central star of a PN.  Our argument is illustrated in Figure~\ref{fig:steps}, which shows the asymptotic reddening as a function of Galactic longitude (Schlafly \& Finkbeiner 2011), plotted for two latitudes, $b$ = 0 (the mid-plane) and $b$ = 0.45 (the latitude of Abell~48).  It is clear that the reddening to Abell 48 is a small fraction of the total reddening in this direction.   The sightline passes close to the intersection of the Galactic Long Bar with the Scutum-Centaurus arm where several major star-forming complexes and massive clusters occur, visible only in the IR (Negueruela et al. 2010, 2012, and references therein).  The heavy interstellar extinction precludes the \emph{optical detection} of any nebula beyond $D$ = 6\,kpc on this sightline.

We can also determine a kinematic distance if we assume the nebula partakes of circular motion around the Galactic centre, albeit with an associated distance ambiguity in this direction. Since Abell\,48 shows no obvious sign of an ISM interaction, any peculiar velocity is not large (cf. TK13).  The systemic velocity, $v_{\rm lsr}$ = +49\,$\pm$ 3\,\kms (see \S\ref{sec:exp_vel}) indicates a kinematic distance of either 3.2 $\pm$ 1.8~kpc or 12 $\pm$ 2.0~kpc, after assuming the IAU standard circular velocity at the solar circle of 220\,\kms.  The uncertainties on the distances are dominated by the velocity dispersions of the respective stellar populations.  However, owing to the heavy extinction along this sightline, we strongly favour the closer distance.

On the assumption that Abell~48 is a normal planetary, we can calculate a distance using its observed properties.  Using the \ha\ surface brightness -- radius ($S_{\rm H\alpha}$--$r$) relation (Frew \& Parker 2006; Frew et al. 2006, 2013c; Frew 2008), and adopting our angular diameter, reddening, and total \ha\ flux, we derive a distance of 1.6\,$\pm$ 0.4 kpc, updated from a preliminary estimate of 2.0\,kpc (Boji\v{c}i\'c et al. 2013).  Adopting this distance,  the Galactic latitude, $b$ of 0.45$^{\circ}$ leads to a $|z|$ height of $\sim$13\,pc, indicating that Abell 48 is almost on the Galactic mid-plane.

\subsection{Derived Nebular Properties}\label{sec:derived_prop}

In this section we use the surrounding ionized nebula as a diagnostic tool to investigate whether the central star is a massive WN star or a post-AGB star of much lower luminosity.  To do this we use the angular diameter, integrated \ha\ flux and the  reddening to determine the nebular size and ionized mass, at distances appropriate for the two contrasting interpretations of the central star.  In particular, the ionized mass is an important reality check of our distance, and therefore all of the nebular and stellar properties that depend on it.
We calculated the ionized mass of the nebula following Hua \& Kwok (1999), who derived the expression: 
\begin{equation} 
M_{ion} = 0.032\, (\epsilon/0.6)^{0.5} \theta^{1.5} D^{2.5} F_{0}({\rm H}\alpha)^{0.5} \, M_{\odot}
\end{equation}
where $\theta$ is in arcmin, $D$ is in kpc, $F_{0}({\rm H}\alpha)$ is the dereddened flux in units of 10$^{-12}$erg\,cm$^{-2}$\,s$^{-1}$, and $\epsilon$ is the  
filling factor.

\begin{table}
\begin{center}
\caption{An abbreviated comparison of quantities, taken from sections 4--6, and a qualitative assessment of the likelihood that Abell~48 a PN or a Population~I ring nebula respectively.}\label{table:ENvsPN}
\begin{tabular}{l c c}
\hline
Parameter 					& PN 	& ~~Ring nebula~~ \\ 
\hline
Distance (kpc) 				& 1.6 	& 12.0 \\
Radius (pc) 					& 0.16 	& 1.2 \\
$M_{\rm ion}/M_{\odot}$ 		& 0.3$\sqrt{\epsilon}$  & 50$\sqrt{\epsilon}$ \\
\hline
Nebular morphology 			& \tick  	& \notick \\
Nebular ionized mass 		& \tick 	& \notick \\
Extinction--distance relation 	& \tick 	& \notick \\
Diagnostic line ratios 			& \tick 	& \notick \\
Nebular abundances 			& \tick 	& \notick \\
Expansion velocity			& \tick 	& \tick \\
Nebular MIR colours			& \tick  	& \notick\\
MIR/radio flux ratio  			& \tick 	& \tick \\
$|z|$ distance from plane~~~~	& \tick 	& \tick \\
\hline
\end{tabular}
\end{center}
\end{table}

We estimate the mass for two different distance estimates, applicable to a PN central star and  a massive WN star.  At 1.6\,kpc, the estimated ionized mass has a sensible value of $M_{\rm ion}$ $\simeq$ 0.3$\sqrt{\epsilon}$~$M_{\odot}$.  This is consistent with other PNe, which typically have ionized masses ranging from 5$\times$10$^{-3}$ to 3 $M_{\odot}$ (FP10).   
For the massive star interpretation, we adopt a distance of 12.0\, kpc.  This leads to a nebular mass, $M_{\rm ion}$ $\simeq$ 50$\sqrt{\epsilon}$~$M_{\odot}$, which even allowing for a very low filling factor, can be ruled out.   
The nebula has  a mean radius of 0.16 pc if  the distance is 1.6\,kpc, typical of a middle-aged PN.  Using  $v_{exp}$ = 40~km~s$^{-1}$ (\S\,\ref{sec:exp_vel}), the dynamical age is $\sim$3900 years with an uncertainty dominated by the distance error (or 30\%).   A summary of parameters for the two contrasting interpretations  is presented in Table \ref{table:ENvsPN}.  In summary, based on the overall body of evidence (see FP10), Abell 48 is clearly a planetary nebula and not a Population I ring nebula on the far side of the bulge.  

\section{A Survey of [WN] Central Stars}\label{sec:WNclass}

The [WR] CSPNe descend from intermediate-mass progenitors, exhibiting a different subclass distribution to the WR stars.  Among the 100+ Galactic [WR] CSPNe, almost all are either [WC] or [WO] stars (Crowther et al. 1998; DePew et al. 2011).  The existence of a [WN] class has long been controversial (Hamann et al. 2003; Werner \& Herwig 2006; FP10; Todt et al. 2010b; MC12).   WN stars are found in the nebulae M\,1-67 and DuRe\,1 (Bertola 1964; Duerbeck \& Reipurth 1990), both formerly classified as PNe, but are now known to be WN8 stars surrounded by CNO-processed ejecta (Cohen \& Barlow 1975; Crawford \& Barlow 1991a,b; Marchenko et al. 2010; Fern\'andez-Mart\'in et al. 2013).   

Since then, the central stars of LMC-N66 (Pe\~na 1995; Pe\~na et al. 2004; Hamann et al. 2003, 2005) and PMR~5 (Morgan, Parker \& Cohen 2003) were classified as possible [WN] stars, and a new [WN/WC] class (Todt et al. 2010a,b, 2012) was introduced to accommodate the spectroscopic characteristics of the ionizing star of PB~8.  More recently, DePew et al. (2011) identified the CSPN of Abell~48 as either a [WN] or [WN/C] star (cf. Wachter et al. 2010) and the CSPN of IC 4663 was found to be a bona fide [WN3] star (MC12).   The CSPN of IC~4663 is rather weak-lined and is likely to be close to its evolutionary transformation into a O(He) star (MC12).  The surface chemistry is dominated by helium, with a modest amount of nitrogen (about a quarter that of Abell~48), no hydrogen, and a relatively low oxygen abundance.

LMC-N66  (SMP\,83) is an unusual, high-excitation, quadrupolar nebula (Pe\~na \& Ruiz 1988; Dopita et al. 1993; Pe\~na et al. 1997) in the Large Magellanic Cloud.  It is ostensibly a PN, as it has a diameter and ionized mass within PN limits, though with a high expansion velocity (Dopita et al. 1985).  The central star is both photometrically variable (McKibben Nail \& Shapley 1955) with an amplitude of $\sim$1 mag, and spectroscopically variable (Torres-Peimbert et al. 1993; Pe\~na 2002; Pe\~na et al. 2008);  Pe\~na (1995) classified it as a [WN4--5] star.  
A good review of this object was provided by Hamann et al. (2003), who investigated several interpretations for it.  All things considered, they concluded that a binary evolution channel is most likely, either a massive star which has lost its surrounding hydrogen through a common-envelope interaction with a less massive companion, or it represents a white dwarf (WD) that is accreting mass from a companion and undergoes quasi-stable nuclear burning.  If this is the case, it may be related to the supersoft X-ray (SSX) sources (Greiner 2000) and their kin, the V~Sge stars (Diaz \& Steiner 1995; Steiner \& Diaz 1998).   The latter scenario would suggest that N66 is a potential Type Ia SN progenitor.   

We note that the unusual quasi-Wolf-Rayet star HD\,45166, which is a close binary (Steiner \& Oliveira 2005; Groh, Oliveira \& Steiner 2008) exhibits a WN5 spectrum, and transient nitrogen-rich WR winds are also seen in  symbiotic novae (Thackeray \& Webster 1974; Nussbaumer 1996) and are probably due to phases of quasi-stable nuclear burning on the WDs in these binary systems.     

Recently, the PN status of PMR~5 was questioned by Werner \& Herwig (2006) and Todt et al (2010b).  We have now re-classified this star from a deep WiFeS spectrum, and revise the classification to WN7b.  We have also found that the surrounding ejecta is composed of strongly CNO-processed material.   A high-resolution VLT \ha\ + \NII\ image showing a flocculent appearance was presented by MC12, showing morphological similarities to RCW~58 (Chu 1982; Stock \& Barlow 2010) and PCG~11 (Cohen, Parker \& Green 2005) which surround WN8h stars.  A fuller account on PMR\,5 will be published separately (Frew et al., in preparation).

\begin{figure*}
\begin{center}
\includegraphics[width=16.cm]{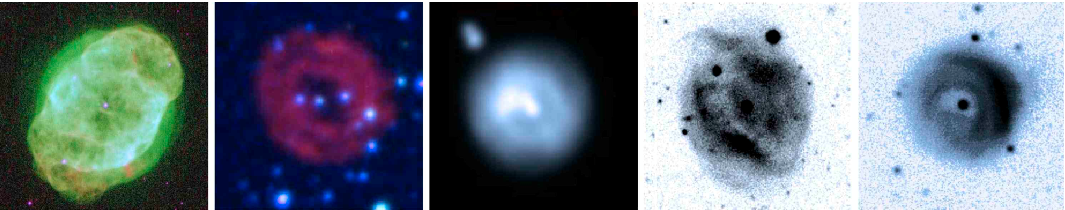}
\caption{A montage of the bona fide Galactic PNe with [WN] and [WN/WC] CSPNe, plus the PNe K~1-27 and LoTr~4 which surround O(He) stars (not to the same scale).  Left to right are IC~4663, Abell~48, PB~8, K~1-27 and LoTr~4.  Abell 48 is a H$\alpha$/SR/B$_{J}$ false-colour composite, while the images of IC~4663, PB~8, K~1-27, LoTr~4 are taken from Hajian et al. (2007), Schwarz et al. (1992), Rauch et al. (1998) and Rauch et al. (1996) respectively.  A colour version of this figure is available in the online journal.  }
\label{WNMontage}
\end{center}
\end{figure*}

Figure~\ref{WNMontage} presents a montage of the three confirmed Galactic PNe with [WN] and [WN/WC] stars, along with the two PNe surrounding O(He) stars, K~1-27 and LoTr~4, included as a morphological comparison.  Table~\ref{comp_parameters} compares the fundamental properties of Abell~48, IC~4663, and PB~8, and their CSPNe.  All three PNe are middle-aged, with PB~8 being the smallest object with the highest \ha\ surface brightness. These values are corrected for reddening, and are derived from the dimensions given in Tylenda et al. (2003).  None of these PNe are particularly luminous, as their absolute $\lambda$5007 magnitudes are 2.6 to 3.8 magnitudes below the bright cut-off of the PN luminosity function (e.g. Ciardullo 2010).  Furthermore, the fact that none of these nebulae are particularly young leads to the question: what are their progenitors?

\begin{table}
\caption{Properties of Abell 48, IC 4663, and PB 8, and their CSPNe.  We include the results for Abell\,48 from TK13 as a comparison.  Data for IC\,4663 and PB\,8 are from MC12 and Todt et al. (2010a) respectively, unless otherwise indicated. }\label{comp_parameters}
\begin{center}
\begin{tabular}{l c c c c c}
\hline
 Object      								& Abell 48$^b$ 	& Abell 48$^c$	& IC 4663    		& PB 8 		\\
\hline
Distance (kpc) 							& 1.6 			& 1.9 			& 3.5 			& 4.2	 	\\  
Radius (pc)								& 0.16 			& 0.2 			& 0.13  			& 0.07$^d$ 	  \\
Nebular age (yr)							& 3900 			& 6500			& 4900  			& 3200$^d$ 	  \\
log\,$S$(\ha)$^a$ 						& $-$2.2 		& ...				& $-$2.2$^d$   	& $-$1.5$^d$   \\
$M_{5007}$								& $-$0.7	 		& ...		 		& $-$1.9$^d$   	& $-$0.7$^d$   \\
\hline
Spectral type  							& [WN4--5]  		& [WN5]  		&  [WN3]   		& [WN6/C7] 	\\
$T_{\star}$ (kK) 							& 71 			& 70 			&  140    		& 52  		\\
$L_{\star}$ ($L_{\odot}$) 					& 5500 			& 6000 			&  4000    		& 6000 		 \\
$v_{\infty}$ (km s$^{-1}$) 					& 1200 			& 1000 			& 1900  			& 1000 		 \\
log\,$R_{t}$ ($R_{\odot}$) 				& 0.8 			& 0.85 			& 1.13 			& 1.43 		  \\
$R_{\star}$ ($R_{\odot}$) 					& 0.49 			& 0.54 			&  0.11   		& 0.96 		  \\
log\,$\dot{M}$ ($M_{\odot}$yr$^{-1}$) 	&$-6.3^{+0.3}_{-0.2}$ & $-$6.4\,$\pm$\,0.2	&   $-7.7^{+0.6}_{-0.2}$   & $-$7.1$^{+0.2}_{-0.1}$    \\
\hline
\end{tabular}
\end{center}
\begin{flushleft}
$^a$Reddening-corrected surface brightness (erg\,cm$^{-2}$s\,$^{-1}$\,sr$^{-1}$); $^b$This~work;  $^c$TK13;  ~$^d$Frew et al. (in preparation).
\end{flushleft}
\end{table}

\section{Siblings, Progenitors and Progeny}\label{sec:siblings}
\subsection{Siblings}\label{sec:A48_siblings}
The surface composition of the CSPN of Abell~48 (see Table~\ref{table:CSPNabund}) is similar to the products of hydrogen-shell burning, i.e. mainly helium with a moderate amount of nitrogen (see also MC12 and TK13), though the nitrogen surface abundance in the CSPN of Abell~48 is to our knowledge, the highest known in a post-AGB star.  If the carbon and oxygen in the progenitor was fully converted to nitrogen via efficient CNO cycling, then the progenitor must have had a substantially supersolar metallicity to explain the high observed nitrogen abundance, but this appears marginally inconsistent with the nebular abundances (\S~\ref{sec:A48_abundances}).  Otherwise, additional mixing and mass-loss processes have occurred.  The existence of  [WN/WC] and [WN] CSPNe implies that there are additional post-AGB evolutionary pathways leading to hydrogen deficiency than previously assumed (De Marco 2002; Werner \& Herwig 2006; Werner 2012; Frew \& Parker 2011, 2012), but at present the full diversity of abundances is not explained by stellar evolution theory.  

Recall there are three primary post-AGB evolutionary scenarios which are thought to produce H-deficient stars, all of which involve a thermal pulse (Iben et al. 1983;  Herwig 2001; Werner \& Herwig 2006), but which differ from each other in the point in the stellar evolutionary track when the thermal pulse occurs.  These are the AGB Final Thermal Pulse (AFTP), the Late Thermal Pulse (LTP), and the Very Late Thermal Pulse (VLTP) scenarios (see Werner \& Herwig 2006, for a review), and the photospheric nitrogen abundance of the star in question is an important diagnostic of the stage during which the final pulse occurs (Werner 2012).

Similar surface abundances to the [WN] stars are also seen in a few, but not all, of the low-gravity, `luminous' He-sdO stars (e.g. Husfeld et al. 1989; Jeffery 2008; Napiwotzki 2008; Heber 2009), none of which are convincingly associated with PNe (M\'endez et al. 1988).  These stars (Heber 2009), alternatively termed sdO(He) stars (Rauch et al. 1998), lie in the same region of the Hertzsprung-Russell (HR) diagram\footnote{The position of the luminous sdO stars in the HR diagram is consistent with post-AGB evolution, while the `compact'  sdOs are likely to derive from a post-extreme horizontal branch (EHB) route (Heber 2009). There are no known PNe around any post-EHB stars (Frew et al. 2010).} as Abell 48.  Indeed, a close analogue is the sdO(He) star LSE~263 (Husfeld et al. 1989) but which is not surrounded by any detectable PN.    
It is feasible that some luminous He-sdO stars are just the lower-mass analogues of the [WN] stars.  If this is correct, the smaller $L/M$ ratios in the He-sdOs explains the absence of any strong wind features, as they are less proximate to the Eddington Limit, and their lower masses explain the absence of any detectable PNe, owing to the long post-AGB timescales of these `lazy' remnants (Sch\"onberner 1983; M\'endez et al. 1988; Bl\"ocker 1995).   
Table~\ref{table:CSPNabund} gives a summary of the photospheric abundances seen in a range of H-normal and H-deficient post-AGB stars, grouped on the basis of their compositions. The top group contains H-normal stars, and the next two groups are helium-rich, greater than about 80\% by mass, but differing in the ratios of N to C\,+\,O.  The following two groups have substantial residual hydrogen, again separated by their N to C ratios, while the lower three groups are those with PG1159 abundances (He\,$\sim$\,C\,$>$\,O), hydrogen-rich or `hybrid' PG1159 compositions (Napiwotzki \& Sch\"onberner 1991), and H/He-deficient compositions, respectively.   It is likely that there are gradual transitions between the various groups, especially of the hydrogen and helium contents (e.g. Rauch \& Kerber 2005), and that the divisions in Table~\ref{table:CSPNabund} are somewhat arbitrary.

\begin{table*}
\caption{Properties and elemental surface abundances (expressed as percentage mass fractions) of a selection of post-AGB stars and pre-WDs.
}  
\label{table:CSPNabund}
\begin{center}
\begin{tabular}{lccccccccccc}
\hline
      Name           		&    SpT   &$T_{\star}$\,(kK)  &log\,$g$\,(cgs)&$X_{\rm H}$&~$X_{\rm He}$~& ~$X_{\rm C}$&~$X_{\rm N}$ & $X_{\rm O}$ & $X_{\rm Ne}$&  PN?  &Reference  \\
\hline
\multicolumn{12}{c}{H-rich composition; ~C + O $>$ N} \\
K 648					&  sdO 				&  	39		&   3.9	&     74      	&   24     	&       0.9    		&     0.001      	&    1.2        		&        ...          	&	y	  & RH02	  \\
NGC 1535				&  O(H)				&     85	 	&   4.8 	&     74       	&    25		&       0.2    		&      0.1        	&    0.8      		&       0.2        	&	y	  &   HB04a, HB11      	\\ 
Sh 2-216				&  DAO				&     95	 	&   6.9	&     95	      	&   4		&       0.1  	     	&     0.1	      		&    	0.2	      		&        ...          	&	y	  &   RZ07	\\
Lo 1					&  O(H)				&     120	&   6.7 	&     69       	&    28		&       0.2   		&      0.1        	&    2.5      		&       0.2	       	&	y	  &   HB04b          	\\ 
\hline
\multicolumn{12}{c}{He-rich composition; ~N $\gg$ C + O} \\
V652 Her             		&    EHe-B           	&   	25	 	&    3.7	&    0.2   	&    99		&     0.01	     	&     0.8           	&   0.04         	&     0.3          	&	n	  & 	JH99    	  	\\
LSE 263              		&    sdO(He)          	&  	70 		& 	4.9	&    $<$2   	&   98           &       0.03          	&     1.6          	&    ...          	 	&        ...          	&	n	  & H89	 \\
Abell 48             			&    [WN4--5]            	&   	71		& $4.9$	&    10		&   85		&       0.3         	&     5.0:           	&     $<$0.6  	&        ...           	&	y	  & 	TK13, this work	  \\
LoTr 4                			&    O(He)          		&    120		& 	5.5	&     12     	&    87       	&     $<$0.07  	&    0.7           	&   $<$0.02       	&     $<$0.1    	&	y	  & 	R13 	\\
PG 1034+001			&    DO				& 	120		&    6.7	&     $<$1.2	&    98    	&       0.02      	&    0.4    		&   0.04     	 	&     0.13		&	n$^{\dagger}$ &	MR12 \\
K 1-27                			&    O(He)          		&  	130	 	& 	6.0	&     4.7     	&    93        	&    $<$0.06       &     1.3  	    	&  $<$0.01    	&    $<$0.5    	&	y	  &	R13 	\\
IC 4663             			&    [WN3]           	&   	140	 	&    6.1	&    $<$2   	&    $>$95   &     $<$0.1     	&     0.8           	&   0.05         	&     0.2          	&	y	  & 	MC12	  \\
\hline
\multicolumn{12}{c}{He-rich composition; ~N $<$ C + O} \\
R CrB					&   G0\,Iab:pe      	&  	6.8		&    0.5 	&       0.001	&     97    	&       1.4         	&     0.3          	&    1.2        		&        ...          	&  	(y)     &   AG00	 \\
BD\,+10\arcdeg 2179	&   EHe-B         		&  	16		&    2.5	&        0.01	&     99    	&       1.7         	&     0.1          	&    0.1        		&      0.1          	&  	n       &   PL11	 \\
BD\,+37\arcdeg 442	&    sdO(He)          	&  	48		&    4.0 	&    $<$0.1 	&     97  	&       2.5          	&     0.3          	&    ...         	 	&        ...          	&  	n       &   BH95, JH10	 \\
HS 2209+8229 			&    O(He)            	&  	110	 	& 	6.0	&    $<$0.6 	&    99      	&    $<$0.01      	&    $<$0.1		&   0.1               	&  $<$0.01	 	&	n	  &	 R13	 	\\
HS 1522+6615  		&    O(He)            	&  	130	 	& 	5.9	&     0.8     	&    98	  	&       1.0           	&    $<$0.01	&   0.13			&  $<$0.03 		&	n	  &	 R13	\\
KPD 0005+5106  		&    DO      			&     200    	& 	6.7	&   $<$2.5  	&    98 		&       1.0   		&     0.3         	&   0.4              	&      0.4     		&	n  	  &    WW10	    \\   
\hline
\multicolumn{12}{c}{He $\geq$ H; ~C $\sim$ N} \\
PB 8                    		&    [WN6/C7]     	&  	52	 	&    4.2	&    40      	&   55    	&       1.3         	&     2.0            	&   1.3   	   	&        ...          	&	y	  & 	T10	  	\\
KS 292                		&    sdO              		&  	75		&    5.0	&    32   	&   65   		&       2.3          	&     1.3          	&    ...          	  	&        ...          	&	n	  & 	R91		\\
\hline
\multicolumn{12}{c}{He $\geq$ H; N $\gg$ C + O} \\
NGC 2392          		&   O6f            		&  	47	 	&   4.0	&     42      	&   57    	&       0.008      	&     0.4      		&    0.01       	&        ...          	&	y	  &	MU12, HB11	\\
HD 49798				&   sdO5.5            	&  	47	 	&   4.3	&     19  	&   78        	&    $<$0.01  	&     2.5	   		&    ...        	     	&        ...          	&	n	  & 	BP97	\\
\hline
\multicolumn{12}{c}{PG 1159 composition (He $\sim$ C + O)} \\
NGC 40                     	&  [WC8]               	&    78          & 	5.0	& $<$2	      	&   43     	&      51               &     ...   	       		&   6                    &    ...                  	&	y	  &   MD07   		   \\
Abell 58                     	&  [WC4]			&     95      	& 	...	& ...		      	&   54 		&      40               &      ...	       		&   5                 	&        ...              &  	y	  &  CK06    \\ 
NGC 6751         			&  [WO4]        		&   135     	& 	6.0	& ...		      	&   54      	&      31             	&      1.5	       	&   15            	&  ...                  	&	y	  &  LK93, KH97		\\
PG 1159-035             	& PG1159            	&    140    	& 	7.0	& $<$2      	&  33   		&      48              	&      0.1           	&   17        	       &       2     	       &	n  	  &  JR07	   \\
RX J2117.1+3412~~ 	&   PG1159			&   163		&     6.6	& ...   		&    39		&    	32		       	&    ...          	  	&     22    		&    ...			&	y	  &  CA07  \\
Lo 4                          		&  [WO]-PG1159  	&   170       	& 	6.0	& ...		      	&   38    	&      54               &      ...	       		&   6                 	&       2               	&	y	  &  WRK10    \\  
\hline
\multicolumn{12}{c}{Hybrid PG 1159 composition (He $\sim$ C $>$ H, O)} \\
IRAS 21282+5050       	& [WC11]h              	&    28   	&  	3.2  & 10             &    43   	&     46               &     $<$0.5        	&    1.0                &  ...  			&   y  	&  WH06      	  \\ 
Abell 43                      	& PG1159h             	&    105   	& 	5.6 	& 24         	&    56     	&     19               	&    0.02         	&   0.2      		&  ...  			&	y	  & RF11  	\\  
HS 2324+3944           	& PG1159h             	&    130     	&  	6.2	& 17   		&    35     	&     42                &      ...              	&    6      		&  ...   			&   n  	& 	DW96  \\ 
\hline
\multicolumn{12}{c}{H- and He-deficient composition (C $\sim$ O)} \\
H\,1504+65				&   PG1159pec		&   200		&     8.0				& ...   		&    $<$1       &   48		       	&    ...          	  	&    48    		&    2		&	n	  &  WR04  \\
\hline
\end{tabular}
\end{center}
\begin{flushleft}
{\scriptsize \emph{References:} ~~ AG00 -- Asplund et al. (2000);  BH95 -- Bauer \& Husfeld (1995);  BP97 -- Bisscheroux et al. (1997);  CA07 -- Corsico et al. (2007);  CK06 -- Clayton et al. (2006);  
DW96 -- Dreizler et al. (1996);  H89 -- Husfeld et al. (1989); HB04a, HB04b --  Herald \& Bianchi (2004a,b); HB11 -- Herald \& Bianchi (2011);  JH99 -- Jeffery, Hill \& Heber (1999);  JR07 -- Jahn et al. (2007);  KH97 -- Koesterke \& Hamann (1997);  LK93 -- Leuenhagen et al. (1993);  MC12 -- Miszalski et al. (2012b); MD07 -- Marcolino et al. (2007);  MR12 -- Mahsereci et al. (2012);  MU12 --  M\'endez et al. (2012);   PL11 -- Pandey \& Lambert (2011);  R91 -- Rauch et al. (1991);   R13 -- Reindl et al. (2013);   RF11 --  Ringat et al. (2011);  RH02 -- Rauch et al. (2002); RW95 -- Rauch \& Werner (1995);  RZ07 -- Rauch et al. (2007);  T10 -- Todt et al. (2010a);  TK13 -- Todt et al. (2013);  WH06 -- Werner \& Herwig (2006); WR04 -- Werner et al. (2004); WW10 -- Wasserman et al. (2010).~~
\emph{Notes:}  $^{\dagger}$PG 1034+001 is the ionizing star of the ``planetary nebula'' Hewett~1, but this is probably just ionized ambient ISM (Chu et al. 2004; Madsen et al. 2006; Frew 2008).
}
\end{flushleft}
\end{table*}

\subsection{Progenitors and Progeny}\label{sec:A48_progenitors}

Werner \& Herwig (2006), Werner (2012) and MC12 have commented on the strong compositional and evolutionary links between the O(He) and [WN] stars, but the overall origin of this group of objects is currently being debated. Only four O(He) stars are known, two with surrounding nebulae (Rauch, Dreizler \& Wolff 1998; Reindl et al. 2013). However, the two stars without observed PNe, HS~1522+6615 and HS~2209+8229, appear to nitrogen poor (see Table~\ref{table:CSPNabund}) and may be evolutionary unrelated to the [WN] stars.  Another example might be Abell~52 (Rauch \& Kerber 2005), which has the \NV$\lambda,\lambda$4604,20 lines in emission, and has detectable hydrogen, with He/H $\approx$ 4 by mass.  Other He-rich stars such as the naked sdO, KS~292 (Rauch et al. 1991), and its PN-shrouded analogue GJJC~1 in M~22 (Harrington \& Paltoglou 1993) have rather more hydrogen, and moderate nitrogen and carbon abundances.  

Rauch et al. (2008) suggested two possible evolutionary scenarios for the formation of the O(He) stars.  They are either the offspring of the R\,CrB stars which are strongly H-deficient cool, variable supergiants, that may derive in turn from low-mass WD mergers (Clayton et al. 2007), or alternatively, they might be post early-AGB or post-RGB stars (Werner 2012).  
Wassermann et al. (2010) found the hot He-rich pre-WD KPD\,0005+5106 to have a similar photospheric abundances to the R\,CrB (Clayton 1996; Asplund et al. 2000) and the extreme-helium (EHe) stars (Pandey et al. 2006; Pandey \& Lambert 2011).  Wassermann et al. (2010) also considered KPD\,0005+5106 and the R\,CrB stars to be WD merger products which form a separate evolutionary post-AGB sequence.  

On theoretical grounds, Saio \& Jeffery (2002) concluded that a merger of a carbon-oxygen WD with a less-massive helium WD is a viable model for the formation of the `majority' R~CrB stars, which are compositionally dominated by helium, and have almost no hydrogen.  More recently, Hema et al. (2012) found that the observed $^{12}$C/$^{13}$C isotopic ratios in the majority R~CrB stars are also consistent with simple nucleosynthesis predictions for such a WD merger (see also Pandey \& Lambert 2011; Jeffery, Karakas \& Saio 2011).  
Yet, the observational evidence for R CrB itself is contradictory (Clayton et al. 2011), as the star is surrounded by a dusty nebula (or fossil PN?) suggesting a post-AGB origin.  There is also a small group of RCB stars with some residual hydrogen and a peculiar abundance pattern, the `minority' RCB stars, but their origin is also a mystery (Rao \& Lambert 2008).   
As noted by TK13, the hydrogen abundance of Abell~48 is $\sim$10\%.  In this respect the CSPN of Abell 48 differs both from the [WN/C] ionizing star of PB~8, and the CSPN of IC~4663 (Todt et al. 2010a; MC12).   An AFTP is unlikely to produce a star with an abundance pattern as seen in Abell~48, as hydrogen abundances are predicted to exceed 15\%, the exact amount depending on the mass of the envelope at the time of the AGB final dredge-up. Moreover, a hybrid PG~1159 composition is expected, whereby the C (+\,O) abundance should greatly exceed the N abundance (Werner \& Herwig 2006).   An LTP scenario is also excluded, as this reduces the nitrogen abundance to only $\sim$0.1\%, and the emission lines attributed to \NIV\ and \NV\ would be expected to be very weak in the optical (Werner et al. 2009).   On the other hand a VLTP model requires that the remaining hydrogen will be completely consumed, with surface abundances less than 10$^{-7}$ by mass, however, the observed CNO mass fractions of Abell~48 are not matched by this scenario.
It is important to recall that the standard LTP scenario (Herwig 2001; Werner \& Herwig 2006) cannot explain the helium-rich abundances seen in these stars.

\subsection{Are Binary Channels Needed?}
So how do the [WN] stars fit in to this framework?  The observed surface abundances appear to require the substantial processing of initial carbon and oxygen to nitrogen, followed by removal of the H-rich envelope, possibly via mass transfer in an Algol-type binary system.  Recall however, that the nebular abundances for Abell 48 and IC 4663 are close to solar (MC12; TK13; this work), so the observed PNe are not simply the stripped atmospheres of these stars.  Furthermore, with the exception of a few objects like V652 Her (Jahn et al. 2007), most EHe (and the R\,CrB) stars differ from the [WN] stars in having photospheric N $<$ C, suggesting  different evolutionary routes for the two groups.  There appears to be a similar dichotomy amongst the O(He) stars, though firm conclusions will only come with a larger sample.  Not withstanding these caveats, a binary post-AGB (or even a post-RGB) evolutionary channel with mass exchange (Iben \& Tutukov 1985), may be required to produce the [WN] stars, rather than an outright merger (cf. MC12).  Guerrero et al. (2013) has suggested an analogous scenario for the formation of the quadrupolar PN, Kn~26 (Jacoby et al. 2010).

\begin{table*}
\caption{Potential evolutionary pathways for post-AGB stars.}\label{table:pathways}
\begin{center}
\begin{tabular}{lll}
\hline
Initial condition     		& Pathway  										 												& Examples 			\\
\hline
\smallskip
No final He flash			&	TP-AGB  $\rightarrow$ O(H) $\rightarrow$ DAO $\rightarrow$ DA	 									&	NGC 1535, NGC 7293, Sh\,2-216	\\
\smallskip
AFTP					&	TP-AGB  $\rightarrow$ Ofc/[WC]  $\rightarrow$ PG1159h  $\rightarrow$ DA	 							&	Abell 43, NGC 7094, Sh\,2-68		\\
\smallskip
LTP						&	post-AGB $\rightarrow$ AGB $\rightarrow$ [WC]  $\rightarrow$  [WO] $\rightarrow$ PG1159  $\rightarrow$ DO $\rightarrow$ DB~~~~~ 	&NGC\,40, PG1159-035, PG1520+525		\\
\smallskip
Weak LTP?				&	post-AGB $\rightarrow$ AGB  $\rightarrow$ [WN/C]h $\rightarrow$ DAO $\rightarrow$ DA	 									&	PB 8		\\
\smallskip
VLTP					&	pre-WD $\rightarrow$ AGB  $\rightarrow$ [WC]  $\rightarrow$  [WO] $\rightarrow$ PG1159  $\rightarrow$ DO $\rightarrow$ DB 	&	Sakurai's object, FG Sge		\\
\hline
\smallskip
Mass transfer / stripping?	&	eAGB $\rightarrow$ He-sdO or [WN] $\rightarrow$ O(He) $\rightarrow$ DO $\rightarrow$ DB	 				&	Abell~48, LMC N66?		\\
\smallskip
He-WD + CO-WD merger~~~~~ &	 R CrB $\rightarrow$ EHe $\rightarrow$ O(He) $\rightarrow$ DO $\rightarrow$ DB &	 R\,CrB, 	KPD 0005+5106?	\\
\smallskip
High-mass progenitor   	&	SAGB $\rightarrow$ PG1159pec/DO? $\rightarrow$ hot DQ $\rightarrow$ DQZ      						&	H1504+65?					\\
\hline
\end{tabular}
\end{center}
\end{table*}

Recently, Herald \& Bianchi (2011) found the unusual CSPN of the Eskimo nebula, NGC\,2392, to show enhanced nitrogen and a deficiency of carbon and oxygen, but with substantial residual hydrogen.  This had been earlier noted by M\'endez (1991) and Pauldrach, Hoffmann \& M\'endez (2004).  This abundance pattern suggests a second dredge-up may have occurred (e.g. Kingsburgh \& Barlow 1994; Marigo et al. 2003), but this is expected only for a narrow range of initial mass, between 3--5 $M_{\odot}$.   However, other evidence suggests that the Eskimo CSPN is derived from a lower-mass progenitor (Pottasch, Bernard-Salas \& Roellig 2008).   Interestingly, M\'endez et al. (2012) suggest that the abundance pattern is the result of truncated evolution due to a close binary interaction, and there is observational evidence (Pottasch et al. 2008; Danehkar et al. 2012) which suggests that its CSPN has an accreting companion, which is detected in hard X-rays (Kastner et al. 2012; Ruiz et al. 2013).  So are the [WN] stars also the product of a binary scenario?

Alternatively, there may be other more exotic post-AGB channels worth considering.  For example, Miller Bertolami et al. (2011) have suggested that diffusion-induced CNO-flashes in a cooling WD, the diffusion-induced nova (DIN) scenario, might apply to PB~8 and the peculiar variable CK~Vul (Hajduk, van Hoof \& Zijlstra 2013, and references therein).  While the surface abundance of nitrogen is  enhanced in the DIN scenario, we deem this to be unlikely for Abell~48, as the requirement of a low-metallicity WD progenitor does not appear to be met.  Furthermore, the nebular ionized mass is typical of post-AGB ejected material, and such a significant mass ejection is not expected from a WD with a cooling age of 10$^6$ to 10$^7$ years (see Miller Bertolami et al. 2011).

Overall, a variety of evolutionary pathways for CSPN evolution is indicated, based on a review of the recent and current literature (e.g. M\'endez 1991; Werner 1992; Rauch et al. 1998; Herwig 2001; Bl\"ocker 2001; De Marco et al. 2002; Miller Bertolami \& Althaus 2006; Werner \& Herwig 2006; Todt et al. 2010a; Werner 2012; Quirion, Fontaine \& Brassard 2012).  Table~\ref{table:pathways} summarizes these pathways.  The first column defines an initial condition for the evolutionary route, the second column gives the route in symbolic form, and the last column gives some candidate examples, primarily drawn from Table~\ref{table:CSPNabund}.
On the basis of the data presented in Table~\ref{table:pathways}, it seems that more theoretical work is required to explain post-AGB stars with an Abell~48-like  abundance pattern, with or without residual surface hydrogen.  It seems the last two proposed evolutionary pathways are the most viable alternatives, either a previously unknown post-AGB evolutionary route, perhaps a post-early AGB pathway, with mass exchange truncating the evolution, or a binary merger of two WDs.

\section{Conclusions and Future Work}
On the basis of an extensive set of new and archival observations, we have undertaken a comprehensive analysis of the the planetary nebula Abell~48 and its CSPN, first classified by Wachter et al. (2010) and DePew et al. (2011).  Based on the presence of very strong \NIV\ lines and moderately strong \NV\ lines at $\lambda\lambda$4604,4620 relative to \NIII\ $\lambda\lambda$4634,4640, we now classify it as [WN4--5] star with little or no hydrogen.   We confirm Abell 48 as a bona fide PN after a detailed comparison of the properties of the nebula and its central star.  If it was a massive star, the distance would be about 12 kpc, and the ionized mass of the nebula would be nearly $\sim$50 $M_{\odot}$, which is much too high even for a Population I ejecta nebula.  Moreover if it were located at or beyond the Galactic Bar, it would suffer much more extinction than observed, and would be optically invisible.  We have estimated a new distance of 1.6~kpc, and an ionized mass of $\sim$0.3$M_{\odot}$, a value typical of other PNe.  Our results are in substantive agreement with the independent study by TK13.

The relationships of the [WN] and [WN/C] stars to other post-AGB stars were also investigated, and we consider it likely that there are two separate channels producing He-rich stars, as distinct from the He- and C- dominated PG1159 stars (see Table~\ref{table:CSPNabund} and Table~\ref{table:pathways} for examples).
Population modelling should be useful in helping to unravel the origin of the [WN] and related objects. Unfortunately these stars are rare ---  only three or four Galactic [WN] and [WN/C] stars and four or five O(He) stars are presently known (Rauch et al. 1998, 2009; Reindl et al. 2012; MC12; TK13, and this work).  The space density of the O(He) stars is expected to be higher than their [WN] progenitors, in a similar vein to the observed ratio of PG~1159 to [WC] stars in a volume-limited sample of PNe (refer to the data of Frew 2008).  This is because the observed space densities are directly related to the time a star spends on the nuclear burning and WD cooling tracks respectively.   

As noted by Frew \& Parker (2011, 2012), the current sample of the so-called weak emission-line stars (WELS; Tylenda et al. 1993;  Gesicki et al. 2006), a seemingly diverse group, may hold several more [WN] and [WN/C] stars awaiting identification on the basis of better spectra.  Indeed, PB~8 was previously identified as a transitional object (M\'endez 1989; Parthasarathy et al. 1998) and IC~4663 as a `WELS' (Weidmann \& Gamen 2011), before their discovery as [WN] stars.   In light of this, there are several more candidate [WN] or [WN/C] stars that we are following up, including PMR~1 (Morgan, Parker \& Russeil 2001), PMR~3 (Parker \& Morgan 2003), PMR~8 (Mabee 2004), PHR\,J1757-1649 (DePew et al. 2011) and Kn\,15 (Kronberger et al. 2012).
The CSPN of M~1-37 (Kudritzki et al. 1997; Hultzsch et al. 2007) is also a possible  [WNL] candidate (see Figure 4 of Kudritzki et al. 1997).

Until more [WN] and O(He) stars are identified, and robust estimates of their space density, scale height and birth rate are established, it is will be difficult to fully understand their relationships to other H-deficient stars, and consequently their formation mechanism(s).  Time will tell if the [WN] CSPNe are the offspring of binary-induced mass exchange, a WD merger, or instead represent the progeny of a previously unknown single-star post-AGB evolutionary pathway.  It is clear that much more work is still to be done.

\section*{Acknowledgements}

We thank the anonymous referee for comments that significantly improved this manuscript.   D.J.F. thanks Macquarie University for a MQ Research Fellowship and I.S.B. is the recipient of an Australian Research Council Super Science Fellowship (project ID FS100100019).  This research has made use of the SIMBAD database and the VizieR service, operated at CDS, Strasbourg, France, and used data from the AAO/UKST \ha\ Survey, produced with the support of the Anglo-Australian Telescope Board and the UK Particle Physics and Astronomy Research Council (now the STFC).  Q.A.P and M.S. acknowledge support from the Australian Astronomical Observatory.

\end{document}